\documentclass[12pt]{article}

\RequirePackage{amsmath,amsthm,amsfonts,amssymb}
\RequirePackage{graphicx}

\usepackage{float}
\usepackage{enumitem}
\usepackage{url}
\usepackage{bbm}
\usepackage{natbib}
\usepackage[nottoc,numbib]{tocbibind}
\usepackage{booktabs,caption,subcaption}
\usepackage[flushleft]{threeparttable}
\usepackage{multirow}
\newcommand{\ra}[1]{\renewcommand{\arraystretch}{#1}}
\usepackage{xr}
\usepackage{color}
\usepackage{tikz}
\usetikzlibrary{patterns}
\usepackage{comment}

\usepackage[margin=1in]{geometry}
\allowdisplaybreaks
\usepackage{setspace}
\newcommand{\ba}{\boldsymbol{a}}
\newcommand{\bA}{\boldsymbol{A}}

\newcommand{\bc}{\boldsymbol{c}}
\newcommand{\bC}{\boldsymbol{C}}
\newcommand{\mCCL}{\mathcal{CCL}}
\newcommand{\mCL}{\mathcal{CL}}

\newcommand{\mD}{\mathcal{D}}
\newcommand{\bg}{\boldsymbol{g}}
\newcommand{\bG}{\boldsymbol{G}}
\newcommand{\bh}{\boldsymbol{h}}
\newcommand{\bH}{\boldsymbol{H}}
\newcommand{\bi}{\boldsymbol{i}}
\newcommand{\bI}{\boldsymbol{I}}

\newcommand{\bJ}{\boldsymbol{J}}
\newcommand{\mN}{\mathcal{N}}

\newcommand{\mP}{\mathcal{P}}
\newcommand{\bs}{\boldsymbol{s}}

\newcommand{\mS}{\mathcal{S}}

\newcommand{\bW}{\boldsymbol{W}}

\newcommand{\mX}{\mathcal{X}}
\newcommand{\by}{\boldsymbol{y}}
\newcommand{\mY}{\mathcal{Y}}
\newcommand{\bz}{\boldsymbol{z}}
\newcommand{\bzero}{\boldsymbol{0}}
\newcommand{\bbeta}{\boldsymbol{\beta}}

\newcommand{\bDelta}{\boldsymbol{\Delta}}

\newcommand{\boeta}{\boldsymbol{\eta}}
\newcommand{\blambda}{\boldsymbol{\lambda}}
\newcommand{\bOmega}{\boldsymbol{\Omega}}
\newcommand{\bPi}{\boldsymbol{\Pi}}
\newcommand{\bPsi}{\boldsymbol{\Psi}}
\newcommand{\bpsi}{\boldsymbol{\psi}}

\newcommand{\btheta}{\boldsymbol{\theta}}

\newtheorem{theorem}{Theorem}

\theoremstyle{definition}

\allowdisplaybreaks

\begin{document}

\begin{singlespace}

\title{\bf Distributed Inference for Spatial Extremes Modeling in High Dimensions}

\author{Emily C. Hector and Brian J. Reich \thanks{This work was supported by grants from the National Science Foundation (DMS2152887, CBET2151651) and the National Institutes of Health (R01ES031651-01). The authors thank Dr. Sankarasubramanian Arumugam of North Carolina State University for providing the streamflow data.}\hspace{.2cm}\\
Department of Statistics, North Carolina State University}
\date{}
\maketitle

\begin{abstract}
Extreme environmental events frequently exhibit spatial and temporal dependence. These data are often modeled using max stable processes (MSPs).  MSPs are computationally prohibitive to fit for as few as a dozen observations, with supposed computation-ally-efficient approaches like the composite likelihood remaining computationally burdensome with a few hundred observations. In this paper, we propose a spatial partitioning approach based on local modeling of subsets of the spatial domain that delivers computationally and statistically efficient inference. Marginal and dependence parameters of the MSP are estimated locally on subsets of observations using censored pairwise composite likelihood, and combined using a modified generalized method of moments procedure. The proposed distributed approach is extended to estimate spatially varying coefficient models to deliver computationally efficient modeling of spatial variation in marginal parameters. We demonstrate consistency and asymptotic normality of estimators, and show empirically that our approach leads to a surprising reduction in bias of parameter estimates over a full data approach. We illustrate the flexibility and practicability of our approach through simulations and the analysis of streamflow data from the U.S. Geological Survey.
\end{abstract}

\noindent%

\end{singlespace}

\vfill

\newpage

\section{Introduction}\label{s:intro}

Despite its immediate practical use and real-world relevance, the modeling of spatial extremes using max-stable processes (MSP) remains theoretically and computationally challenging in high spatial dimensions. The main technical challenge lies in adequately capturing spatial dependence using low-dimensional marginal projections of the joint distribution of the spatial extreme outcomes while controlling the computational burden of the analysis as the dimension of these marginal distributions increases \citep{Huser-Wadsworth-2022}. To balance these two fundamental necessities, we propose a data partitioning approach that leverages recent advances in divide-and-conquer techniques for dependent outcomes and delivers three new tools for analysis of spatial extremes: (i) a censored pairwise likelihood approach for analysis of spatial extremes when the MSP model is only valid for outcomes above a threshold, (ii) a computationally and statistically efficient divide-and-conquer meta-estimator that integrates censored pairwise likelihood information from all parts of the spatial domain, and (iii) a flexible analytic toolbox for spatially-varying coefficient MSP models in high dimensions.

The MSP models pointwise maxima over infinitely many independent realizations of a spatial process, and provides a general and flexible class of models for spatial extremes through \cite{deHaan}'s spectral representation. The extremal tail dependence is specified by a general exponential measure for which many models have been proposed, such as \cite{Smith-1990, Tawn, Schlather, Kabluchko-Schlather-deHaan, Buishand-etal, Wadsworth-Tawn-2012}. Theoretical assumptions for MSPs are frequently difficult to satisfy in practice: extreme events are by definition rare, so that there are often not enough replicates to justify the theoretical approximation to maxima over infinitely many observations. When pointwise maxima are taken over a small to moderate number of replicates, the MSP yields a poor fit \citep{Huang-etal}. A viable solution is the use of censored likelihoods to model the dependence between observations above a threshold \citep{Thibaud-Mutzner-Davison, Huser-Davison-2014}. The censored likelihood approach uses the partial information available on the extremal coefficients from points below the threshold without requiring strong modeling assumptions. This approach has primarily been used to model univariate/marginal points above thresholds using the generalized Pareto distribution \citep{Ledford-Tawn, Smith-etal, Bortot-etal, Coles, Wadsworth-Tawn-2014, Thibaud-Opitz}. \cite{Huser-Davison-2014} extended the univariate thresholding approach to bivariate thresholding for pairwise censored likelihood inference. As we discuss below, the computational cost of pairwise censored likelihood methods remains high, and the analysis of extreme values on large spatial domains persists as an open problem.

The analytic form of (censored) MSP densities is computationally intractable for all but trivially small spatial fields. Only a few models for the exponential measure have closed-form bi- or tri-variate densities \citep{Schlather, Kabluchko-Schlather-deHaan, Wadsworth-Tawn-2012}, with higher-order densities typically impossibly complex. The primary difficulty lies in computing the exploding number of partial derivatives of the exponential measure. For example, the Brown-Resnick process has $d$th order density consisting of $B_d$ terms with $B_d$ the $d$th Bell number \citep{Wadsworth-Tawn-2014}. This has led to the predominant use of the composite likelihood (CL) \citep{Lindsay, Varin-Reid-Firth}. The core philosophy of the CL approach is to construct marginal likelihoods on subsets of data and integrate them using working independence assumptions; thus the CL is not a proper likelihood, but a product of proper likelihoods. The most widely used form of the CL, the pairwise CL, approximates the likelihood by the product of bivariate likelihoods. The seminal work of \cite{Padoan-Ribatet-Sisson} cemented the pairwise CL as a practical and versatile method for inference with MSPs by formally defining the procedure and examining its theoretical properties. Since then, the pairwise and triplewise CL have played a prominent role in computationally attractive methodological developments for inference with MSPs \citep{Genton-Ma-Sang, Davison-Gholamrezaee, Huser-Davison-2013, Sang-Genton, Castruccio-etal, Huser-Genton}.

The pairwise CL is attractive because it offers a trade-off between statistical efficiency and computational speed. Moreover, the maximum CL estimator is consistent and asymptotically normal under mild regularity conditions \citep{Padoan-Ribatet-Sisson}. The pairwise CL still suffers from loss of efficiency that is particularly evident for large dimensions \citep{Huser-Davison-Genton}. In addition, for $d$ observation locations, the pairwise CL evaluates bivariate densities for ${d \choose 2}=O(d^2)$ pairs of observations, which may become computationally burdensome for $d \gtrsim 10^2$. CL estimation of MSP parameters based on all pairs of observations also suffers from finite-sample bias \citep{Sang-Genton, Wadsworth-2015, Castruccio-etal}. 

Spatially-varying coefficient models that allow marginal parameters to vary by observation location are essential for modeling the spatial distribution of extreme events. While spatially-varying coefficient MSP model fitting tools are available in R packages \citep{SpatialExtremes}, to our knowledge these have not been investigated in the CL literature except under a working independence model \citep{sass2021flexible}, presumably due to the tremendous computational burden of estimating a large number of parameters with pairwise CL. An alternative strategy capable of handling spatially-varying coefficients that makes use of dependence between some pairs of observations is highly desirable.

We propose a new local model building approach for spatial extreme value analysis that constructs censored pairwise CL on subsets of spatial observations and integrates these dependent CLs using a modified Generalized Method of Moments (GMM) objective function \citep{Hansen}. The resulting integrated censored pairwise CL estimator is statistically and computationally efficient, and exhibits a surprisingly reduced bias compared to the standard censored pairwise CL estimator. Our approach hinges on two key observations for the construction of the GMM weight matrix: (i) the optimal choice of the GMM matrix is the sample covariance matrix of the pairwise composite score functions, which yields an estimator with variance at least as small as any other estimator constructed from the same pairwise score functions; (ii) this weight matrix introduces finite sample bias in the integrated estimator by accounting for dependence between all subsets of spatial observations. To trade-off between the desire for both optimal efficiency and reduced bias, we propose a new weighting matrix that strikes a balance between these two goals, and show how the resulting estimator can be estimated using a computationally appealing meta-estimator implemented in the MapReduce paradigm. We extend this approach to spatially-varying marginal regression models for added modeling flexibility. We show through simulations that our approach's tremendous computational advantage enables MSP inference with potentially thousands of spatially dependent extreme value observations, a heretofore unattainable goal.

The rest of the paper is organized as follows. We review the MSP construction and existing approaches in Section \ref{s:MSP}. In Section \ref{s:GMM}, we describe the proposed data partitioning, local model construction, and censored pairwise CL integration approach. We extend the proposed framework to spatially-varying coefficient models in Section \ref{s:SVCM}. The finite sample performance of the proposed estimator is investigated through simulations in Section \ref{s:simulations}. An analysis of flood frequency data from the U.S. Geological Survey is presented in Section \ref{s:data}. Theoretical conditions and derivations, additional data analysis results and an R package are provided in the online supplementary materials.

\section{Problem Set-up}\label{s:MSP}

\subsection{The Max-Stable Process}\label{ss:MSP:background}

Let $\mS \subset \mathbb{R}^2$ a spatial domain and $Y_r(\bs)$ the extreme value at location $\bs \in \mS$ for replicate $r\in\{1,\ldots,m\}$. Assume that $Y(\bs)$ is the block-maximum, i.e., $Y(\bs) = \max\{Y_1(\bs),\ldots,Y_m(\bs)\}$. Considering the joint distribution of the point-wise maximum of the $m$ realizations at all locations in $\mS$ gives the random field ${\cal Y} = \{Y(\bs) ; \bs \in \mS\}$. Under certain regularity conditions, ${\cal Y}$ can be well approximated by a max-stable process (MSP) for large $m$. See also the excellent reviews of \cite{Ribatet-book} and \cite{Davison-Huser-Thibaud}.

Assuming the process is max-stable, then the marginal distribution of $Y(\bs)$ is the generalized extreme value (GEV) distribution $\mbox{GEV}\{\mu(\bs), \sigma(\bs), \xi(\bs)\}$, where $\mu(\bs)$ is the location, $\sigma(\bs)>0$ is the scale, and $\xi(\bs)$ is the shape. The GEV parameters can be allowed to vary spatially to capture local differences in the magnitude of extreme values. The MSP can be written equivalently as 
\begin{align*}
Y(\bs)=\mu(\bs)+ \frac{\sigma(\bs)}{\xi(\bs)} \left[ X(\bs)^{\xi(\bs)}-1 \right],
\end{align*}
where ${\cal X}=\{X(\bs);\bs\in\mS\}$ is a MSP with unit Fr\'{e}chet marginal distributions, $X(\bs) \sim \mbox{GEV}(1,1,1)$. The three GEV parameters explain spatial variation in the marginal distribution, whereas the spatial dependence of ${\cal X}$ explains residual variation. For example, if $Y(\bs)$ is the annual maximum (i.e., $m=365$) of daily precipitation at $\bs$, then $\mu(\bs)$, $\sigma(\bs)$ and $\xi(\bs)$ determine the distribution of the annual maximum across years at $\bs$, whereas the spatial dependence of ${\cal X}$ determines the likelihood that two locations will simultaneously experience an above average rainfall amount in a given year. 

The finite-dimensional distribution function of any MSP at $d$ locations $\mD=\{\bs_1,\ldots,\bs_d\}$ has the form $\mbox{Prob}\{ X(\bs_i) < x_i, i=1, \ldots, d \}=\exp \{-V(x_1,\ldots,x_d) \}$ for some exponential measure $V$ that satisfies 
$V(\lambda x_1,\ldots,\lambda x_d) = V(x_1,\ldots, x_d)/\lambda$ for any $x_1,\ldots,x_d,\lambda>0.$ Under the assumption that $X(\mD)=\{X(\bs_1), \ldots, X(\bs_d)\}$ has unit Fr\'{e}chet marginal distributions, then the exponential measure must satisfy $V(x_1,\ldots, x_d) = 1/x_j$ if $x_i=\infty$ for all $i\ne j$. Of the many possibilities for the exponential measure, we choose the Brown-Resnick model \citep{Brown-Resnick, Kabluchko-Schlather-deHaan} because it gives a stationary process and provides flexibility in modeling the smoothness of ${\cal X}$ across space. The exponential measure that defines the joint distribution function of the pair $X(\bs_i)$ and $X(\bs_j)$ is 
\begin{equation*}
V(x_i,x_j; \alpha, \phi) = \frac{1}{x_i}\Phi\left\{\frac{a_{ij}}{2} - \frac{1}{a_{ij}}\log\left(\frac{x_i}{x_j}\right)\right\} +
\frac{1}{x_j}\Phi\left\{\frac{a_{ij}}{2} - \frac{1}{a_{ij}}\log\left(\frac{x_j}{x_i}\right)\right\},
\end{equation*}
where $\Phi$ is the standard normal distribution function, $a_{ij} = \left\{2\gamma(\bs_i-\bs_j)\right\}^{1/2}$ and $\gamma$ is a semivariogram. Following \cite{Huser-Davison-2013}, we use the isotropic semivariogram $\gamma(\bs_i-\bs_j) = (||\bs_i-\bs_j||/\phi)^\alpha$ defined by spatial range $\phi>0$ and smoothness $\alpha\in[0,2]$. Our objectives are to estimate the GEV parameters $\mu(\bs)$, $\sigma(\bs)$ and $\xi(\bs)$ and the spatial dependence parameters $\phi$ and $\alpha$ when the number $d$ of observation locations is large.

\subsection{Model and Existing Approaches} \label{ss:MSP:model}

We consider the setting with $n$ independent replicates of $Y(\mD)$ denoted by $Y_1(\mD), \ldots, Y_n(\mD)$, where $Y_i(\mD)=\{Y_i(\bs_1), \ldots, Y_i(\bs_d) \}^\top$ and $\mD=\{\bs_j\}_{j=1}^d$ the set of observation locations. Correspondingly, we have $n$ independent replicates of $X(\mD)$ denoted by $X_1(\mD), \ldots, X_n(\mD)$, where $X_i(\mD)=\{X_i(\bs_1), \ldots, X_i(\bs_d)\}^\top$ is related to $Y_i(\mD)$ through
\begin{align}
Y_i(\bs_j)=\mu_i(\bs_j)+ \frac{\sigma_i(\bs_j)}{\xi_i(\bs_j)} \left[ X_i(\bs_j)^{\xi_i(\bs_j)}-1 \right],
\label{e:Y-X-rel}
\end{align}
for $i=1, \ldots, n$, $j=1, \ldots, d$. Let $\bz_i(\cdot)$ be $q$ explanatory variables observed at spatial locations $\bs_1, \ldots, \bs_d$ for replicate $i\in \{1, \ldots, n\}$. For $\bz_{i1}, \bz_{i2}, \bz_{i3} \subseteq \bz_i$ of respective dimensions $q_1,q_2,q_3$, we posit the model
\begin{align*}
\mu_i(\bs; \bbeta_1)&=\bz_{i1}(\bs)^\top \bbeta_1,\quad 
\sigma_i(\bs; \bbeta_2)=\exp(\bz_{i2}(\bs)^\top \bbeta_2),\quad
\xi_i(\bs; \bbeta_3)=\bz_{i3}(\bs)^\top \bbeta_3.
\end{align*}
A more flexible spatially-varying coefficient model is introduced in Section \ref{s:SVCM}.  Let $\bbeta=(\bbeta^\top_1,\bbeta^\top_2,\bbeta^\top_3)^\top$. To facilitate estimation of $\alpha,\phi$, we propose the reparametrization $\omega=\log\{ \alpha/(2-\alpha)\}$, $\zeta=\log(\phi)$, or equivalently $\alpha=2\exp(\omega)/\{1+\exp(\omega)\}$, $\phi=\exp(\zeta)$, and let $\btheta=(\omega, \zeta, \bbeta^\top)^\top \in \mathbb{R}^p$. The analytic goal is to estimate and make inference on $\btheta$. 

When $d=2$ or $3$, estimation and inference on $\btheta$ using maximum likelihood is feasible since the full likelihood of the Brown-Resnick process has a closed form following \cite{Huser-Davison-2013} and \cite{Ribatet-book}; see also the online supplementary materials. For $d>3$, however, the full likelihood generally becomes analytically challenging to derive and computationally burdensome to evaluate. For general MSPs, \cite{Castruccio-etal} stated that full likelihood inference seemed limited to $d=12$ or 13 by then-current technologies. More recently, \cite{Huser-etal} proposed an expectation-maximization algorithm for full likelihood inference for $d>13$, and illustrated the performance of their algorithm for the Brown-Resnick process for dimensions up to $d=20$. Their approach, however, remains computationally prohibitive for large $d$ due to the evaluation of multivariate Gaussian probabilities, with computation time of $19.8$ hours when $d=20$.

Composite likelihood (CL) \citep{Lindsay, Varin-Reid-Firth} has therefore become the method of choice to overcome the computational burden of full likelihood inference. Denote $\{ \mD_1,\ldots, \mD_K\}$ a partition of $\mD$ such that $\mD=\cup_{k=1}^K \mD_k$ and $\mD_j \cap \mD_k = \varnothing$ for $j \neq k$. Denoting $y_i(\mD)=\{y_i(\bs_j) \}_{j=1}^d$, the log composite likelihood assumes working independence between observations in different sets $\mD_k$ and takes the form
\begin{align*}
\mCL\{ \btheta; y_i(\mD), i=1, \ldots, n\}= \log \prod_{i=1}^n \prod_{k=1}^K f\{ y_i(\mD_k); \btheta \}= \sum_{i=1}^n \sum_{k=1}^K \log f\{ y_i(\mD_k); \btheta \},
\end{align*}
where $f\{ y_i(\mD_k); \btheta\}$ is the multivariate marginal density of $y_i(\mD_k)=\{y_i(\bs): \bs \in \mD_k \}$. The pairwise CL has cardinality $|\mD_k|=2$ of $\mD_k$, $K={d \choose 2}$, and has been widely used in spatial extreme value analysis; see for example the review of \cite{Davison-Padoan-Ribatet} and references in Section \ref{s:intro}. While including all pairs of observations leads to decreased variance, empirical studies have shown that it also leads to increased finite sample bias. This phenomenon appears to have been first observed by \cite{Sang-Genton}, who proposed a tapering weight to improve efficiency of the pairwise CL but noted substantial bias of estimators, in particular of the smoothness $\alpha$, for both untapered and tapered pairwise CL. \cite{Wadsworth-2015} investigated the source of the bias in large dimensions $d$ when incorporating information on occurrence times. \cite{Castruccio-etal} and \cite{Huser-Davison-Genton} conducted thorough empirical studies of the bias with various CL-based methods and generally illustrated that bias increases as $d$ is large relative to $n$. 

In addition to increased bias, the pairwise CL remains computationally burdensome when $d$ is large. This difficulty stems from the need to evaluate analytically complex bivariate likelihoods at all pairs of observations. The pairwise CL is therefore not scalable to large $d$ and alternative strategies are required. In Section \ref{s:GMM}, we propose a spatial partitioning approach resulting in the computationally efficient evaluation of pairwise CL on low dimensional subsets of the spatial domain.

\section{A Spatial Partitioning Approach} \label{s:GMM}

\subsection{Partitioning the Spatial Domain} \label{ss:GMM:partition}

\begin{figure}
\centering
\includegraphics[width=0.75\textwidth]{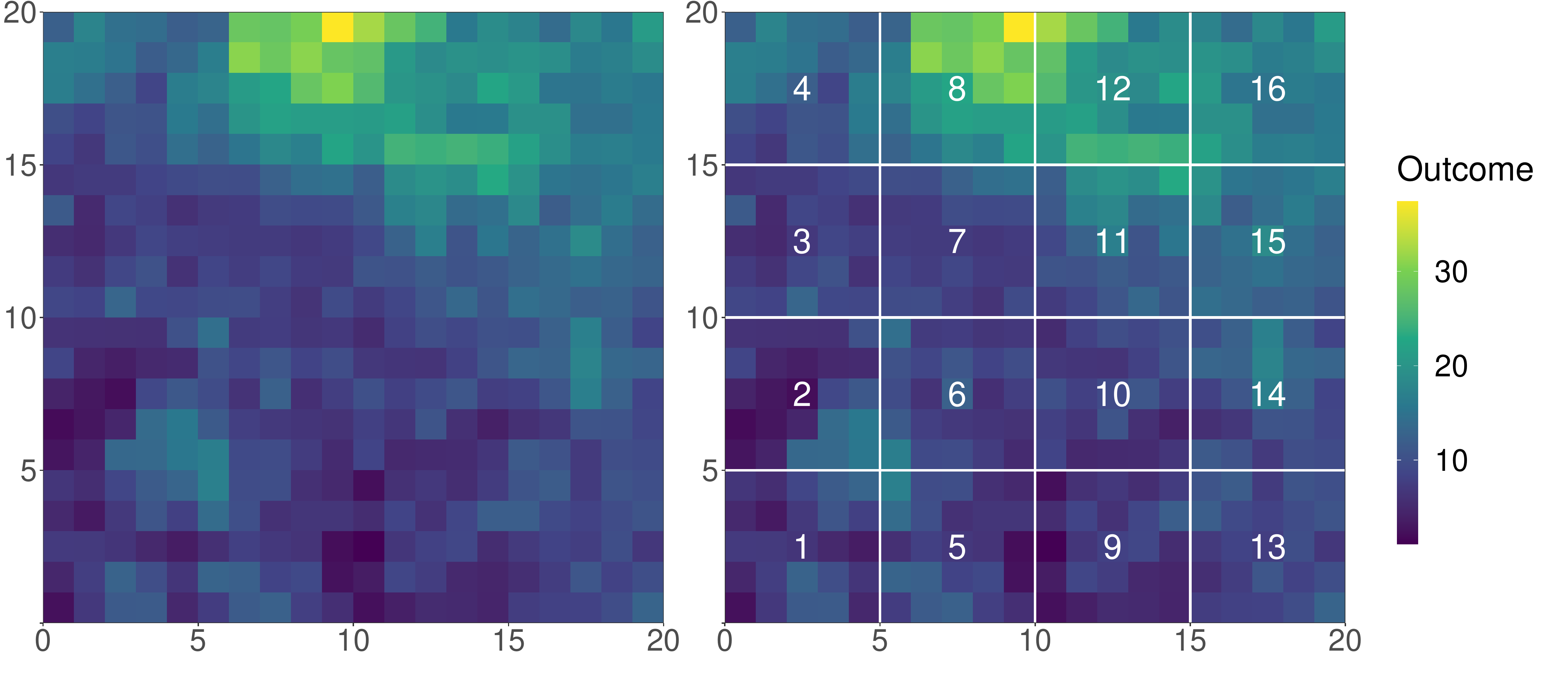}
\caption{Example partition of outcomes on a 2D spatial domain: on the left, observation on a spatial domain of size $20\times 20$; on the right, partitions of size $d_k=25$.}
\label{f:ex_partitions}
\end{figure}

Following the spirit of the CL, we propose a partition of the spatial domain $\mD$ into $K$ disjoint regions $\mD_1, \ldots, \mD_K$ such that $\cup_{k=1}^K \mD_k=\mD$ and denote by $d_k$ the number of observation locations in $\mD_k$. To facilitate estimation of $\btheta$ in each subset $\mD_k$, we partition $\mD$ such that $d_k$ is relatively small, e.g. $d_k=25$; see for example Figure \ref{f:ex_partitions}, where squares represent observation locations. The literature is rich with methods for choosing partitions for Gaussian processes: see for examples \cite{Knorr-Held-Raszer, Kim-etal, Sang-etal, Anderson-etal, Heaton-etal-2017} and the review of \cite{Heaton-etal-2019}. With MSP, due to the nature of the range parameter $\phi$, we generally recommend partitioning $\mD$ into regions of similar size $d_k$ based on nearest locations as in Figure \ref{f:ex_partitions}, with $d_k$ sufficiently large so as to allow for a range of distances between locations in $\mD_k$. Note that many of the spatial partitioning approaches reviewed in \cite{Heaton-etal-2019} assume independence between spatial subsets $\mD_k$, which we do not. 

The local likelihood approach for threshold exceedances of \cite{Castro-Camilo-Huser} resembles the first step of our approach, although the models bear substantial differences and the authors focus on dependence parameter estimation rather than joint-modeling of marginal and dependence parameters.

\subsection{Local Likelihood Specification} \label{ss:GMM:likelihood}

Let $k\in \{1, \ldots, K\}$. Since $d_k$ may still be large enough to render full likelihood estimation of $\btheta$ in $\mD_k$ intractable, we propose to estimate the parameters of interest in subset $\mD_k$ using the pairwise CL approach. Inference with pairwise CL for max-stable processes has been ubiquitous since the seminal paper of \cite{Padoan-Ribatet-Sisson}, but this approach assumes the MSP is an appropriate model for $Y_i(\bs)$, $i=1, \ldots, n$, which may not hold in practice if the block maxima were taken over blocks with small size $m$. Following \cite{Beirlant-etal}, \cite{Huser-Davison-2014} observed that the MSP defined in Section \ref{ss:MSP:background} also models extremes of individual observations, and used a censored likelihood approach for bivariate extremes in the composite likelihood framework to overcome this difficulty.

Inspired by their approach, consider two locations $\bs_1, \bs_2 \in \mD_k$ and denote $y_{ij}=y_i(\bs_j)$, $x_{ij}=x_i(\bs_j)$, $j=1,2$. Let $u_1=u(\bs_1), u_2=u(\bs_2)$ be sufficiently high thresholds such that $f( y_{i1}, y_{i2}; \btheta)$ is a valid model for $y_{i1}, y_{i2} \in \{y_i(\bs): \bs \in \mD_k\}$ when $y_{i1}>u_1,y_{i2}>u_2$, $y_{i1}\neq y_{i2}$, $i=1, \ldots, n$, with $f( y_{i1}, y_{i1}; \btheta)= f(x_{i1}, x_{i2}; \alpha, \phi) J(y_{i1}; \bbeta)J(y_{i2}; \bbeta)$ the bivariate max-stable density obtained from the MSP defined in \eqref{ss:MSP:background}, and the Jacobians $J(y_{i1}; \bbeta)$, $J(y_{i2}; \bbeta)$ and $f(x_{i1}, x_{i2}; \alpha, \phi)=\partial^2 \exp \{ \allowbreak -V(x_{i1}, x_{i2})\}/(\partial x_{i1} \partial x_{i2})$ given in the online supplementary materials. Let $u_{ij}=\{1+\xi_i(\bs_j)(u_j-\mu_i(\bs_j)/\sigma_i(\bs_j)\}^{1/\xi_i(\bs_j)}$, $j=1,2$. The likelihood contribution $g(y_{i1}, y_{i2}; \btheta, u_1, u_2)$ of the pair $(y_{i1},y_{i2})$ for $\bs_1, \bs_2 \in \mD_k$ is
\begin{align*}
\begin{cases}
f(x_{i1}, x_{i2}; \alpha, \phi) J(y_{i1}; \bbeta)J(y_{i2}; \bbeta) , & y_{i1}>u_1,y_{i2}>u_2,\\
\left[ \frac{\partial}{\partial x_{i1}} \exp \left\{ -V(x_{i1}, u_{i2}; \alpha, \phi) \right\} \right] J(y_{i1}; \bbeta) ,& y_{i1}>u_1, y_{i2} \leq u_2,\\
\left[ \frac{\partial}{\partial x_{i2}} \exp \left\{ -V(u_{i1}, x_{i2}; \alpha, \phi) \right\} \right] J(y_{i2}; \bbeta) ,& y_{i1}\leq u_1, y_{i2}>u_2,\\
\exp \left\{ -V(u_{i1},u_{i2}; \alpha, \phi) \right\},& y_{i1} \leq u_1, y_{i2} \leq u_2.
\end{cases}
\end{align*}
Using the censored likelihood pairs, the log censored composite likelihood (CCL) in $\mD_k$ takes the form
\begin{align}
\mCCL_k ( \btheta; u )&=\frac{1}{n} \sum \limits_{i=1}^n \sum \limits_{\mP_k} \log g(y_{i1},y_{i2}; \btheta, u),
\label{e:CCL}
\end{align}
where $\mP_k=\{(\bs_1, \bs_2): \bs_1, \bs_2 \in \mD_k\}$. Clearly, letting $u_1,u_2 \rightarrow -\infty$ recovers the uncensored composite likelihood,
\begin{align*}
\mCL_k(\btheta)&=\sum \limits_{i=1}^n \sum \limits_{\mP_k} \left\{ \log f(x_{i1},x_{i2}; \alpha, \phi) + \log J(y_{i1}; \bbeta) + \log J(y_{i2}; \bbeta) \right\}.
\end{align*}
Based on the log CCL in region $\mD_k$, we obtain the censored composite score function:
\begin{align}
\bPsi_k(\btheta)&=\frac{1}{n} \sum \limits_{i=1}^n \bpsi_{ik}(\btheta)=\frac{1}{n} \left( \frac{\partial}{\partial \omega}, \frac{\partial}{\partial \zeta}, \frac{\partial}{\partial \bbeta_1}, \frac{\partial}{\partial \bbeta_2}, \frac{\partial}{\partial \bbeta_3} \right)^\top \mCCL_k(\btheta) \in \mathbb{R}^p,
\label{e:CCL-score}
\end{align}
with specific form given in the online supplementary materials. Solving $\bPsi_k(\btheta)=\boldsymbol{0}$ yields the maximum censored composite likelihood estimator (MCCLE) $\widehat{\btheta}_k=(\widehat{\omega}_k, \widehat{\zeta}_k, \widehat{\bbeta}_{1k}, \widehat{\bbeta}_{2k}, \widehat{\bbeta}_{3k})^\top$ of $\btheta$. We denote by $\bi_k(\btheta)= E_{\btheta} [\nabla_{\btheta} \{ \bPsi_k(\btheta) \}]$ and $\bc_k(\btheta)=Var_{\btheta} \{ \sqrt{n} \bPsi_k(\btheta) \}$ the sensitivity and variability matrices, respectively, of $\bPsi_k(\btheta)$. 

Let $\btheta_0=(\omega_0, \zeta_0, \bbeta^\top_0)^\top$ the true value of $\btheta$ such that $E_{\btheta} \{ \bPsi_k(\btheta)\}$ is uniquely zero at $\btheta_0$. Consistency and asymptotic normality of the MCCLE $\widehat{\btheta}_k$ are formalized in Theorem \ref{t:MCCLE-asy}.
\begin{theorem}
\label{t:MCCLE-asy}
Under mild regularity conditions \ref{c:1} in Appendix \ref{a:c},
\begin{align*}
\sqrt{n} \left( \widehat{\btheta}_k - \btheta_0 \right) \stackrel{d}{\rightarrow} \mN \left\{ \bzero, \bi^\top_k (\btheta_0) \bc^{-1}_k(\btheta_0) \bi_k(\btheta_0) \right\}.
\end{align*}
\end{theorem}
The proof follows from standard estimating function theory \citep{Lindsay, Godambe-1991, Varin-Reid-Firth, Padoan-Ribatet-Sisson, Huser-Davison-2014}.

\subsection{Censored Composite Likelihood integration} \label{ss:GMM:integration}

Suppose we have successfully obtained the $K$ MCCLEs $\{\widehat{\btheta}_k \}_{k=1}^K$ for regions $\mD_k$, $k=1, \ldots, K$. We now wish to integrate these local estimators into one unified estimator of $\btheta$ over all $K$ regions. An efficient model integration procedure should leverage the dependence between the $K$ MCCLEs, but this dependence is difficult to estimate directly because we do not have replicates of these estimators. Alternative bootstrap-type procedures to estimate this dependence are computationally costly. To overcome this difficulty, we integrate the censored composite score functions in equation \eqref{e:CCL-score} rather the MCCLEs. 

Define the stacking operation $\{\ba_k\}_{k=1}^K=(\ba^\top_1, \ldots, \ba_K)^\top \in \mathbb{R}^{\sum_{k=1}^K D_k}$ and $\{\bA_k\}_{k=1}^K=(\bA^\top_1, \ldots, \bA^\top_K)^\top \in \mathbb{R}^{\sum_{k=1}^K D_{1k} \times D_2}$ for vectors $\ba_k \in \mathbb{R}^{D_k}$ and matrices $\bA_k \in \mathbb{R}^{D_{1k}\times D_2}$. Define the stacked censored composite likelihood kernel and score functions as $\bpsi_{i,all}(\btheta)=\{ \bpsi_{ik}(\btheta)\}_{k=1}^K \in \mathbb{R}^{Kp}$ and $\bPsi_{all}(\btheta)=\left\{ \bPsi_k(\btheta) \right\}_{k=1}^K \in \mathbb{R}^{pK}$ respectively. Denote the sensitivity and variability matrices of $\bPsi_{all}(\btheta)$ by
\begin{align*}
\bi(\btheta)= \{\bi_k(\btheta)\}_{k=1}^K \in \mathbb{R}^{Kp \times p}, \quad \bc(\btheta)=Var_{\btheta} \{\sqrt{n} \bPsi_{all}(\btheta)\} \in \mathbb{R}^{Kp \times Kp},
\end{align*}
respectively. A key insight is that the stacked censored composite score function $\bPsi_{all}(\btheta)$ over-identifies $\btheta$: there are more estimating equations than there are dimensions on $\btheta$. This leads to a natural use of \cite{Hansen}'s generalized method of moments (GMM), which minimizes a quadratic form of the over-identifying moment conditions:
\begin{align}
\widehat{\btheta}_{GMM}&=\arg \min \limits_{\btheta} n\bPsi^\top_{all}(\btheta) \bW \bPsi_{all}(\btheta)=\arg \min \limits_{\btheta} n\sum \limits_{k,k'=1}^K \bPsi_k(\btheta) \left( \bW \right)_{k,k'} \bPsi_{k'}(\btheta),
\label{e:GMM-quad}
\end{align}
where $( \bW )_{k,k'}$ denotes the rows and columns of $\bW$ corresponding to subsets $\mD_k$ and $\mD_{k'}$ respectively, for any positive semi-definite weight matrix $\bW$. This approach has been successfully employed by others \citep{Bai-Song-Raghu, Hector-Song-JASA} although never with a MSP or censored (composite) likelihood, and has connections to weighted composite likelihood \citep{LeCessie-vanHouwelingen, Nott-Ryden, Kuk, Joe-Lee, Zhao-Joe, Sang-Genton, Castruccio-etal}. Under mild regularity conditions \citep{Newey-McFadden}, $\widehat{\btheta}_{GMM}$ is a consistent estimator of $\btheta$ and asymptotically normally distributed as $n \rightarrow \infty$:
\begin{align*}
\sqrt{n} \left( \widehat{\btheta}_{GMM}-\btheta_0 \right) \stackrel{d}{\rightarrow} \mathcal{N} \left\{0, \bOmega(\btheta_0) \bc(\btheta_0) \bOmega^\top(\btheta_0) \right\},
\end{align*}
where $\bOmega(\btheta_0)=-\{ \bi^\top(\btheta_0) \bW \bi(\btheta_0)\}^{-1} \bi^\top(\btheta_0) \bW$. From the presence of $\bc(\btheta)$ in the asymptotic variance of $\widehat{\btheta}_{GMM}$, dependence between the $K$ spatial subsets $\{ \mD_k \}_{k=1}^K$ is incorporated in the evaluation of the estimator's uncertainty. Thus, the GMM estimator is not evaluated under working independence assumptions, and the quantification of the uncertainty of $\widehat{\btheta}_{GMM}$ is robust to the form of the between-subset dependence. 

Following \cite{Hansen}, the most efficient choice of $\bW$ is clearly $\bc^{-1}(\btheta_0)$, which minimizes the diagonal of $\bOmega(\btheta_0) \bc(\btheta_0) \bOmega^\top(\btheta_0)$. This choice is equivalent to using all the dependence between spatial subsets $\{\mD_k\}_{k=1}^K$. It is well known, however, that CL estimation of MSP model parameters based on all pairs of observations suffers from finite-sample bias and under-estimation of the true variance \citep{Sang-Genton, Wadsworth-2015, Castruccio-etal}. When $d$ is large, e.g. $d \gtrsim 100$, estimation of $\bc(\btheta_0)$ based on the full sample covariance matrix $\bC(\btheta)=(1/n)\sum_{i=1}^n \bpsi_{i,all}(\btheta) \bpsi^\top_{i,all}(\btheta)$ may introduce bias into $\widehat{\btheta}_{GMM}$. To overcome this difficulty, we use the fact that, jointly, $\sqrt{n} ( \widehat{\btheta}_k -\btheta_0 )_{k=1}^K \stackrel{d}{\rightarrow} \mathcal{N} ( \boldsymbol{0}, \bs^\top(\btheta_0) \bc^{-1}(\btheta_0) \bs(\btheta_0) )$ \citep{Hector-Song-JMLR}, and so, marginally,
\begin{align*}
\sqrt{n} \left( \widehat{\btheta}_k - \btheta_0 \right) \stackrel{d}{\rightarrow} \mathcal{N} \left[ \boldsymbol{0}, \bi^\top_k(\btheta_0) \left\{ \bc^{-1}(\btheta_0) \right\}_{k,k} \bi_k(\btheta_0) \right].
\end{align*}
This motivates our choice of a weight matrix $\bW$ that mitigates the effect of covariance between subsets on the finite-sample bias of the GMM estimator. We propose $ \bW(\btheta)=\mbox{diag} \{ \bW_k(\btheta) \}_{k=1}^K$, with $\bW_k(\btheta)= \{ \bC^{-1}(\btheta) \}_{k,k} \in \mathbb{R}^{p\times p}$ to obtain the GMM estimator
\begin{align}
\widehat{\btheta}^\star_{GMM} &= \arg \min \limits_{\btheta} n\sum \limits_{k=1}^K \bPsi^\top_k(\btheta) \bW_k(\btheta) \bPsi_k(\btheta).
\label{e:GMM-quad-opt}
\end{align}
Consistency and asymptotic normality of the GMM estimator in equation \eqref{e:GMM-quad-opt} are established in Theorem \ref{GMM-asy}.
\begin{theorem}
\label{GMM-asy}
Under regularity conditions \ref{c:1} and \ref{c:2} in Appendix \ref{a:c}, the proposed GMM estimator $\widehat{\btheta}^\star_{GMM}$ in equation \eqref{e:GMM-quad-opt} satisfies
\begin{align*}
\sqrt{n} ( \widehat{\btheta}^\star_{GMM} - \btheta_0 ) \stackrel{d}{\rightarrow} \mathcal{N} [ \boldsymbol{0}, 
\{ \bh(\btheta_0) \}^{-1}
\bg(\btheta_0)
\{ \bh^\top(\btheta_0)
\}^{-1}
],
\end{align*}
as $n \rightarrow \infty$, where
\begin{align*}
\bh(\btheta)&=\sum_{k=1}^K \bi^\top_k(\btheta) \left\{ \bc^{-1}(\btheta) \right\}_{k,k} \bi_k(\btheta)\\
\bg(\btheta)&=\sum \limits_{k,k'=1}^K 
\bi^\top_k(\btheta) \left\{ \bc^{-1}(\btheta) \right\}_{k,k} 
\left\{ \bc(\btheta) \right\}_{k,k'}
\left\{ \bc^{-1}(\btheta) \right\}_{k',k'} \bi_{k'}(\btheta).
\end{align*}
\end{theorem}

The proof follows immediately from \cite{Hansen}. The proposed GMM estimator in equation \eqref{e:GMM-quad-opt} thus possesses required statistical properties for inference while benefiting from a substantial reduction in computation time.

\subsection{Implementation: a Meta-Estimator} \label{ss:GMM:covariance}

The iterative minimization in \eqref{e:GMM-quad-opt} remains computationally burdensome for large $d$ because the censored composite score function of $d_k$ pairs must be evaluated at each iteration of the minimization. Fortunately, this iterative procedure may be altogether bypassed through the closed-form meta-estimator derived by \cite{Hector-Song-JASA}:
\begin{align}
\widehat{\btheta}_m&=\left\{ \sum \limits_{k=1}^K \bI^\top_k(\widehat{\btheta}_c ) \bW_k(\widehat{\btheta}_c) \bI_k ( \widehat{\btheta}_c ) \right\}^{-1} \sum \limits_{k=1}^K \bI^\top_k( \widehat{\btheta}_c ) \bW_k ( \widehat{\btheta}_c ) \bI_k ( \widehat{\btheta}_c ) \widehat{\btheta}_k
\label{e:meta}
\end{align}
where $\bI_k(\btheta)=\nabla_{\btheta} \bPsi_k(\btheta)$ denotes the sample sensitivity matrix, and $\widehat{\btheta}_c$ is a suitable consistent estimator of $\btheta$ specified as follows. The estimation of $\bI_k(\btheta)$ and $\bW_k(\btheta)$ by $\bI_k(\widehat{\btheta}_c)$ and $\bW_k(\widehat{\btheta}_c)$, respectively, requires careful consideration. These matrices may be estimated by plugging in the MCCLEs, i.e., using $\bI_k(\widehat{\btheta}_k)$ and
\begin{align*}
\left\{ \left( \frac{1}{n} \sum \limits_{i=1}^n \left\{ \bpsi_{ik} (\widehat{\btheta}_k) \right\}_{k=1}^K \left[\left\{ \bpsi_{ik} (\widehat{\btheta}_k) \right\}_{k=1}^K\right]^\top \right)^{-1} \right\}_{k,k},
\end{align*}
but these estimators may have high variability depending on the performance of the MCCLEs in each subset. A better estimator can be constructed from the average of the MCCLEs: $\widehat{\btheta}_c=(1/K) \sum \limits_{k=1}^K \widehat{\btheta}_k$.
This leads to the following distributed procedure:
\begin{enumerate}[itemsep=0em, leftmargin=*]
\item Partition the spatial domain $\mD$ into $K$ disjoint regions $\mD_1, \ldots, \mD_K$.
\item For $k=1, \ldots, K$, estimate $\widehat{\btheta}_k$ in subset $\mD_k$ using the censored composite likelihood in \eqref{e:CCL}. This step can be performed in parallel on $K$ nodes to accelerate computation.
\item Compute the average of the MCCLEs, $\widehat{\btheta}_c=\sum_{k=1}^K \widehat{\btheta}_k/K$, on the main computing node.
\item For $k=1, \ldots, K$, evaluate and return $\bpsi_{ik}(\widehat{\btheta}_c)$ and $\bI_k(\widehat{\btheta}_c)$ to the main computing node. This step can be performed in parallel on $K$ nodes to accelerate computation.
\item Form $\bpsi_{i,all}(\widehat{\btheta}_c)=\{ \bpsi_{ik}(\widehat{\btheta}_c) \}_{k=1}^K$ and compute $\bC(\widehat{\btheta}_c)=(1/n)\sum_{i=1}^n \bpsi_{i,all} (\widehat{\btheta}_c) \bpsi^\top_{i,all}(\widehat{\btheta}_c)$, $\bW_k(\widehat{\btheta}_c)=\{\bC^{-1}(\widehat{\btheta}_c)\}_{k,k}$, and $\widehat{\btheta}_m$ in \eqref{e:meta}.
\end{enumerate}
This distributed approach to estimation of $\btheta$ requires two rounds of communication between distributed nodes and the main computing node: the first to return $\widehat{\btheta}_k$, and the second to return $\bpsi_{ik}(\widehat{\btheta}_c)$ and $\bI_k(\widehat{\btheta}_c)$. The derivative $\bI_k(\btheta)$ can be estimated as the sum of the sample covariance of bivariate censored score functions for each pair of observations in $\mD_k$. This results in a flexible and computationally efficient procedure. Inferential properties of the estimator $\widehat{\btheta}_m$ in \eqref{e:meta} are shown in Theorem \ref{t:meta-asy}.

\begin{theorem}
\label{t:meta-asy}
Under conditions \ref{c:1} and \ref{c:2} in Appendix \ref{a:c}, the proposed estimator $\widehat{\btheta}_m$ in equation \eqref{e:meta} is consistent and asymptotically normally distributed as $n \rightarrow \infty$:
\begin{align*}
\sqrt{n} \left( \widehat{\btheta}_m- \btheta_0 \right) \stackrel{d}{\rightarrow} \mathcal{N} \left[ \bzero, \left\{ \bh(\btheta_0)\right\}^{-1}
\bg(\btheta_0)
\left\{ \bh^\top(\btheta_0)\right\}^{-1}
\right],
\end{align*}
where the asymptotic covariance of $\widehat{\btheta}_m$ in equation \eqref{e:meta} can be consistently estimated by $\bJ^{-1}(\widehat{\btheta}_c)=n^{-1}\{ \bH(\widehat{\btheta}_c) \}^{-1} \allowbreak \bG(\widehat{\btheta}_c) \{ \bH^\top(\widehat{\btheta}_c) \}^{-1}$ with
\begin{equation}
\begin{split}
\bH(\btheta)&=\sum_{k=1}^K \bI^\top_k(\btheta) \bW_k \bI_k(\btheta)\\
\bG(\btheta)&=\sum \limits_{k,k'=1}^K 
\bI^\top_k (\btheta) \bW^{-1}_k(\btheta)
\left[ \bC(\btheta) \right]_{k,k'} 
\bW^{-1}_{k'}(\btheta) \bI_{k'}(\btheta).
\label{e:m-dist}
\end{split}
\end{equation}
\end{theorem}
The proof of Theorem \ref{t:meta-asy} is a special case of the proofs given in the online supplementary materials for the spatially-varying coefficient model (see Section \ref{s:SVCM}). Moreover, it follows easily from equation \eqref{e:m-dist} that $\bJ(\btheta)$ can also be computed in a distributed fashion using the quantities returned from the distributed nodes at the second round of communication. 

Denote $\widehat{\btheta}_m=(\widehat{\omega}_m, \widehat{\zeta}_m, \widehat{\bbeta}^\top_m)^\top$. Let $\alpha_0=2\exp(\omega_0)/\{1+\exp(\omega_0)\}$, $\phi_0=\exp(\zeta_0)$, and $\bDelta \in \mathbb{R}^{p\times p}$ a matrix with $(\alpha_0/\{1+\exp(\omega_0)\}, \phi_0, 1, \ldots, 1)^\top$ on the diagonal and $0$'s elsewhere. If desired, asymptotic normality of $(\widehat{\alpha}_m, \widehat{\phi}_m, \widehat{\bbeta}^\top_m)^\top$, where  $\widehat{\alpha}_m=2\exp(\widehat{\omega}_m)/\{1+\exp(\widehat{\omega}_m)\}$, $\widehat{\phi}_m=\exp(\widehat{\zeta}_m)$, is obtained through the Delta method:
\begin{align*}
\sqrt{n} \left\{ \left(
\widehat{\alpha}_m, \widehat{\phi}_m, \widehat{\bbeta}^\top_m
\right)^\top -
\left( 
\alpha_0, \phi_0, \bbeta^\top_0
\right)^\top \right\} \stackrel{d}{\rightarrow} \mN \left[ \bzero, \bDelta^\top \left\{ \bh(\btheta_0)\right\}^{-1}
\bg(\btheta_0)
\left\{ \bh^\top(\btheta_0)\right\}^{-1} \bDelta \right].
\end{align*}

\section{Extension to Spatially-Varying Coefficients}\label{s:SVCM}

\subsection{Local Model and Likelihood Specification}\label{ss:SVCM:model}

When the spatial variation in $\mu_i(\bs)$ and/or $\sigma_i(\bs)$ is of interest, we propose a spatial varying coefficient model \citep{Hastie-Tibshirani, Serban} for added modeling flexibility. We prefer the varying coefficient model over nonparametric kernel smoothing for its ability to fit in our divide-and-conquer framework; see \cite{Davison-Ramesh} for a univariate (i.e., non-spatial) nonparametric kernel smoothing approach. We describe this model for $\mu_i(\bs)$, with a similar description for $\log \{\sigma_i(\bs)\}$ omitted for brevity. See \cite{Waller-etal} for a review of approaches for spatially varying coefficient models. 

Suppose $\mu_i(\cdot)$ depends on $z_{i1,t}(\cdot)$, the $t$\textsuperscript{th} covariate in $\bz_{i1}(\cdot) \in \mathbb{R}^{q_1}$ for replicate $i$, through some unknown function $b_{t,\mu}(\cdot)$, $t=1, \ldots, q_1$: $\mu_i(\bs)=\sum_{t=1}^{q_1} z_{i1,t}(\bs) b_{t,\mu}(\bs)$. Let $\{\phi_{jt,k}(\cdot)\}_{j,k=1}^{\infty,K}$ be radial basis functions of the functional space to which $b_{t,\mu}(\cdot)$ belongs, $t=1, \ldots, q_1$ \citep{Ruppert-Wand-Carroll}. Within each spatial subset $\mD_k$, we approximate $b_{t,\mu}(\mD_k)$ by a finite linear combination of the basis functions, i.e., $b_{t,\mu}(\mD_k) \approx \sum_{j=1}^{J_{1tk}} \eta_{1jtk} \phi_{jt,k}(\mD_k)$, where $J_{1tk}$ is the number of basis functions for the $t$\textsuperscript{th} covariate function and $\boeta_{1k}=\{ \eta_{1jtk} \}_{j,t=1}^{J_{1tk},q_1}$ is the unknown $p_{\boeta_{1k}}$-dimensional parameter of interest, $p_{\boeta_{1k}}=\sum_{t=1}^{q_1} J_{1tk}$ fixed. Defining $\widetilde{z}_{i1,jt}(\bs)=z_{i1,t}(\bs) \phi_{jt,k}(\bs)$ for $j=1, \ldots, J_{1tk}$, $t=1, \ldots, q_1$, $\bs \in \mD_k$, and substituting into the mean model yields $\mu_i( \bs) \approx \sum_{t=1}^{q_1} \sum_{j=1}^{J_{1tk}} \widetilde{z}_{i1,jt}(\bs) \eta_{1jtk}$ for $\bs \in \mD_k$.

Suppose we propose a similar varying coefficient model expansion for $\log \{ \sigma_i(\bs)\}$ with parameter vector $\boeta_{2k}=\{\eta_{2jtk}\}_{j,t=1}^{J_{2tk},q_2} \in \mathbb{R}^{p_{\boeta_{2k}}}$, where $J_{2tk}$ is the number of basis functions for the $t$\textsuperscript{th} covariate function in the finite linear approximation of $\log \{\sigma_i(\cdot)\}$ in $\mD_k$ and $p_{\boeta_{2k}}=\sum_{t=1}^{q_2} J_{2tk}$ is fixed, $k=1, \ldots, K$. Let $\boeta_k=(\boeta^\top_{1k}, \boeta^\top_{2k})^\top \in \mathbb{R}^{p_k}$, with $p_k = p_{\boeta_{1k}} + p_{\boeta_{2k}}$ fixed. Due to the difficulty in estimating $\xi(\bs)$, we maintain the model $\xi_i(\bs; \bbeta_3)=\bz_{i3}(\bs)^\top \bbeta_3$ proposed in Section \ref{ss:MSP:model}. The MCCLE of $\btheta_k=(\omega, \zeta, \boeta^\top_k, \bbeta^\top_3)^\top$ in subset $\mD_k$ can be computed as in Section \ref{ss:GMM:likelihood} and is denoted by $\widehat{\btheta}_k=(\widehat{\omega}_k, \widehat{\zeta}_k, \widehat{\boeta}^\top_k, \widehat{\bbeta}^\top_{3k})^\top \in \mathbb{R}^{2+p_k+q_3}$, with $\widehat{\boeta}^\top_k= (\widehat{\boeta}^\top_{1k}, \widehat{\boeta}^\top_{2k})$. Note that we are not assuming that $\boeta_{1k} \equiv \boeta_1$, $\boeta_{2k} \equiv \boeta_2$: we model these marginal parameters separately for each subset to retain the spatial variation of the relationship between $\mu_i(\bs)$, $\sigma_i(\bs)$ and $\bz_{i1}(\bs)$, $\bz_{i2}(\bs)$. This results in heterogeneous marginal parameters. On the other hand, $\omega$, $\zeta$ and $\bbeta_3$ are assumed homogeneous across all subsets. 

Let $\widehat{\omega}_c=(1/K)\sum_{k=1}^K \widehat{\omega}_k$, $\widehat{\zeta}_c=(1/K)\sum_{k=1}^K \widehat{\zeta}_k$, $\widehat{\bbeta}_{3c}=(1/K)\sum_{k=1}^K \widehat{\bbeta}_{3k}$, $\widehat{\boeta}_c=( \widehat{\boeta}^\top_1, \ldots, \allowbreak \widehat{\boeta}^\top_K )^\top \in \mathbb{R}^p$ with $p=\sum_{k=1}^K p_k$, and $\boeta=(\boeta^\top_1, \ldots, \boeta^\top_K )^\top$. Under conditions \ref{c:1} in Appendix \ref{a:c}, $\widehat{\btheta}_c=( \widehat{\omega}_c, \widehat{\zeta}_c, \widehat{\boeta}^\top_c, \widehat{\bbeta}^\top_{3c} ) \in \mathbb{R}^{2+ p + q_3}$ is a consistent estimator of $\btheta=(\omega, \zeta, \boeta^\top, \bbeta^\top_3)^\top$.

\subsection{Integration Procedure}\label{ss:SVCM:integration}

The goal of the integration procedure is to update the $K$ estimators $\widehat{\boeta}_k$, $k=1, \ldots, K$, for the heterogeneous parameters and to combine the $K$ estimators $(\widehat{\omega}_k, \widehat{\zeta}_k, \widehat{\bbeta}^\top_{3k})^\top$, $k=1, \ldots, K$, for the homogeneous parameters while leveraging dependence between subsets as in Sections \ref{ss:GMM:integration} and \ref{ss:GMM:covariance}. This will yield an integrated estimator of $\btheta$ that retains the spatial variation of the relationship between $\mu_i(\bs)$, $\sigma_i(\bs)$ and $\bz_{i1}(\bs)$, $\bz_{i2}(\bs)$. This integration can be derived with modification of the framework developed in Section \ref{s:GMM}. Denote
\begin{align*}
\widetilde{\bPsi}_{1k}(\btheta_k)&=\frac{1}{n} \sum \limits_{i=1}^n \widetilde{\bpsi}_{i1k}(\btheta_k)=\frac{1}{n} \left( \frac{\partial}{\partial \omega}, \frac{\partial}{\partial \zeta} \right)^\top \mCCL_k(\btheta) \in \mathbb{R}^2,\\
\widetilde{\bPsi}_{2k}(\btheta_k)&=\frac{1}{n} \sum \limits_{i=1}^n \widetilde{\bpsi}_{i2k}(\btheta_k)=\frac{1}{n} \left( \frac{\partial}{\partial \boeta_{1k}}, \frac{\partial}{\partial \boeta_{2k}} \right)^\top \mCCL_k(\btheta) \in \mathbb{R}^{p_k},\\
\widetilde{\bPsi}_{3k}(\btheta_k)&=\frac{1}{n} \sum \limits_{i=1}^n \widetilde{\bpsi}_{i3k}(\btheta_k)=\frac{1}{n} \frac{\partial}{\partial \bbeta_3} \mCCL_k(\btheta) \in \mathbb{R}^{q_3}.
\end{align*}
Let $\widetilde{\bpsi}_{ik}(\btheta_k)=\{ \widetilde{\bpsi}_{ijk}(\btheta_k) \}_{j=1}^3$, $\widetilde{\bpsi}_{i,all} (\btheta) = \{ \widetilde{\bpsi}_{ik} (\btheta_k) \}_{k=1}^K$, $i=1, \ldots, n$ and $\widetilde{\bPsi}_k(\btheta_k) = \{ \widetilde{\bPsi}_{jk}(\btheta_k) \}_{j=1}^3$, $k=1, \ldots, K$. Let $\widetilde{\bPsi}_{all}(\btheta)=\{ \widetilde{\bPsi}_k(\btheta_k)\}_{k=1}^K$. Define the estimated sample covariance matrix of $\widetilde{\bpsi}_{i,all}(\btheta)$ as $\widetilde{\bC}(\btheta)=(1/n) \sum_{i=1}^n \widetilde{\bpsi}_{i,all}(\btheta) \widetilde{\bpsi}^\top_{i,all}(\btheta) \in \mathbb{R}^{(2K+p+q_3K) \times (2K+p+q_3K)}$. 

Let $\lambda_1, \lambda_2 \geq 0$ denote two tuning parameters, $\blambda_{\eta_j} \in \mathbb{R}^{(2+ p + q_3)\times (2+ p + q_3)}$ denote a diagonal matrix with $\lambda_j$'s in the positions for $\boeta_j$ and 0 elsewhere, $j=1,2$. We define two sensitivity matrices: $\bI_k(\btheta_k)= \nabla_{\btheta_k} \widetilde{\bPsi}^\top_k(\btheta_k) \in \mathbb{R}^{(2+ p_k+q_3) \times (2+p_k+q_3)}$ and $\widetilde{\bI}_k(\btheta)= \nabla_{\btheta} \widetilde{\bPsi}^\top_k(\btheta_k) \in \mathbb{R}^{ (2+p_k+q_3) \times (2+ p+q_3)}$, $k=1, \ldots, K$. By construction of $\btheta$, $\widetilde{\bI}_k(\btheta_k)$ is obtained from $\bI_k(\btheta)$ by adding rows of $0$'s for parameters in $\btheta$ that are not in $\btheta_k$. The varying-coefficient model meta-estimator is given by
\begin{align}
\widehat{\btheta}_{Vm}&= \left\{ \sum \limits_{k=1}^K \widetilde{\bPi}_k(\widehat{\btheta}_c) + \blambda_{\boeta_1} + \blambda_{\boeta_2} \right\}^{-1} \sum \limits_{k=1}^K \bPi_k(\widehat{\btheta}_c) \widehat{\btheta}_k,
\label{e:meta-VCM}
\end{align}
with $\widetilde{\bPi}_k(\btheta) = \widetilde{\bI}^\top_k(\btheta) \{ \widetilde{\bC}^{-1}(\btheta) \}_{k,k} \widetilde{\bI}_k(\btheta)$ and $\bPi_k(\btheta) = \widetilde{\bI}^\top_k(\btheta) \{ \widetilde{\bC}^{-1}(\btheta) \}_{k,k} \bI_k(\btheta_k)$.

Let $\widetilde{\bi}_k=E_{\btheta} [ \nabla_{\btheta} \{ \widetilde{\bPsi}_k(\btheta_k)\}]$ and $\widetilde{\bc}(\btheta)=Var_{\btheta} \{ \sqrt{n} \widetilde{\bPsi}_{all}(\btheta)\}$. Denote by $\btheta_0$ the unique $\bzero$ of $E_{\btheta} \{ \bPsi_{all}(\btheta)\}$, the true value of $\btheta$. 

\begin{theorem}
\label{t:VCM-asy}
When $\lambda_1, \lambda_2 \rightarrow 0$ as $n \rightarrow \infty$ and conditions \ref{c:1} and \ref{c:2} in Appendix \ref{a:c} with $\bPsi_k(\btheta)$, $\bPsi_{all}(\btheta)$, $\bi_k(\btheta)$, $\bc(\btheta)$ replaced with $\widetilde{\bPsi}_k(\btheta_k)$, $\widetilde{\bPsi}_{all}(\btheta)$, $\widetilde{\bi}_k(\btheta)$ and $\widetilde{\bc}(\btheta)$ are satisfied, $\widehat{\btheta}_{Vm}$ in equation \eqref{e:meta-VCM} is a consistent estimator of $\btheta_0$ and asymptotically normally distributed, with asymptotic covariance consistently estimated by $\bJ^{-1}(\widehat{\btheta}_c) = n^{-1} \{ \bH(\widehat{\btheta}_c) \}^{-1} \allowbreak \bG(\widehat{\btheta}_c) \{ \bH^\top(\widehat{\btheta}_c) \}^{-1}$ with
\begin{equation}
\begin{split}
\bH(\btheta)&=\sum \limits_{k=1}^K \widetilde{\bPi}_k(\btheta) + \blambda_{\boeta_1} + \blambda_{\boeta_2}\\
\bG(\btheta)&=\sum \limits_{k,k'=1}^K 
\widetilde{\bI}^\top_k (\btheta) \left\{ \widetilde{\bC}^{-1}(\btheta) \right\}_{k,k} 
\left\{ \widetilde{\bC}(\btheta) \right\}_{k,k'} 
\left\{ \widetilde{\bC}^{-1}(\btheta) \right\}_{k',k'}
\widetilde{\bI}_{k'}(\btheta).
\label{e:Vm-dist}
\end{split}
\end{equation}
\end{theorem}
The proof of consistency and asymptotic normality is given in the online supplementary materials. Since $\lambda_1, \lambda_2 \rightarrow 0$ as $n \rightarrow \infty$, we could replace $\bH(\btheta)$ with $\sum_{k=1}^K \widetilde{\bPi}_k(\btheta)$: asymptotically, these two matrices are equivalent. In practice, accounting for the smoothing induced by $\lambda_1, \lambda_2$ with $\bH(\btheta)$ will yield more precise estimation of the variance of $\widehat{\btheta}_{Vm}$ in finite samples.

Denote $\widehat{\btheta}_{Vm}=(\widehat{\omega}_{Vm}, \widehat{\zeta}_{Vm}, \widehat{\boeta}^\top_{Vm}, \widehat{\bbeta}^\top_{3Vm})^\top$. The asymptotic covariance of $(\widehat{\alpha}_{Vm}, \widehat{\phi}_{Vm}, \allowbreak  \widehat{\boeta}^\top_{Vm}, \widehat{\bbeta}^\top_{3Vm})^\top$, with $\widehat{\alpha}_{Vm}=2\exp(\widehat{\omega}_{Vm})/\{1+\exp(\widehat{\omega}_{Vm})\}$ and $\widehat{\phi}_{Vm}=\exp(\widehat{\zeta}_{Vm})$, can be recovered using the Delta method as in Section \ref{ss:GMM:covariance}. Let $( \widehat{\btheta}_{Vm} )_{\eta_{1jtk}}$ and $( \widehat{\btheta}_{Vm} )_{\eta_{2jtk}}$ denote the elements of $\widehat{\btheta}_{Vm}$ corresponding to parameters $\eta_{1jtk}$ and $\eta_{2jtk}$ respectively. Estimates of $\mu_i(\bs)$ and $\sigma_i(\bs)$ are recovered through
\begin{align*}
\widehat{\mu}_i(\bs)&= \sum \limits_{t=1}^{q_1} \sum \limits_{j=1}^{J_{1tk}} \widetilde{z}_{i1,jt}(\bs) \left( \widehat{\btheta}_{Vm} \right)_{\eta_{1jtk}} , \quad \log \left\{ \widehat{\sigma}_i(\bs) \right\}= \sum \limits_{t=1}^{q_2} \sum \limits_{j=1}^{J_{2tk}} \widetilde{z}_{i2,jt}(\bs) \left( \widehat{\btheta}_{Vm} \right)_{\eta_{2jtk}}, \quad \bs \in \mD_k.
\end{align*}
To reconstruct estimates of the standard errors of $\mu_i(\bs)$ and $\sigma_i(\bs)$, we let $\{ \bJ^{-1} (\widehat{\btheta}_c) \}_{\boeta_1}$ and $\{ \bJ^{-1} (\widehat{\btheta}_c) \}_{\boeta_2}$ denote the submatrices of $\bJ^{-1}(\widehat{\btheta}_c)$ corresponding to parameters $\{ \boeta_{1k} \}_{k=1}^K$ and $\{ \boeta_{2k} \}_{k=1}^K$ respectively. Denote
\begin{alignat*}{4}
\widetilde{\bz}_{i1}(\bs) &= \{\widetilde{z}_{i1,jt}(\bs) \}_{j,t=1}^{J_{1tk},q_1} \in \mathbb{R}^{p_{\boeta_{1k}}},\quad &\widetilde{\bz}_{i2}(\bs) &= \{\widetilde{z}_{i2,jt}(\bs) \}_{j,t=1}^{J_{2tk},q_2} \in \mathbb{R}^{p_{\boeta_{2k}}},\quad
\bs \in \mD_k\\
\widetilde{\bz}_{i1}(\mD_k) &= \{ \widetilde{\bz}_{i1}(\bs)^\top \}_{\bs \in \mD_k} \in \mathbb{R}^{d_k \times p_{\boeta_{1k}}}, \quad
&\widetilde{\bz}_{i2}(\mD_k) &= \{ \widetilde{\bz}_{i2}(\bs)^\top \}_{\bs \in \mD_k} \in \mathbb{R}^{d_k \times p_{\boeta_{2k}}} \\
\widetilde{\bz}_{i1}(\mD)&=\mbox{diag} \left\{ \widetilde{\bz}_{i1}(\mD_k) \right\}_{k=1}^K \in \mathbb{R}^{d \times \sum_{k=1}^K p_{\boeta_{1k}}}, \quad 
&\widetilde{\bz}_{i2}(\mD)&=\mbox{diag} \left\{ \widetilde{\bz}_{i2}(\mD_k) \right\}_{k=1}^K \in \mathbb{R}^{d \times \sum_{k=1}^K p_{\boeta_{2k}} }.
\end{alignat*}
The square roots of the diagonal of
\begin{align*}
\widehat{Var} \left\{ \widehat{\mu}_i(\mD) \right\}&=\widetilde{\bz}_{i1}(\mD) \left\{ \bJ^{-1}(\widehat{\btheta}_c) \right\}_{\boeta_1} \widetilde{\bz}_{i1}(\mD)^\top ,\\
\widehat{Var}\left[ \log \left\{ \widehat{\sigma}_i(\mD) \right\} \right]&=\widetilde{\bz}_{i2}(\mD) \left\{ \bJ^{-1}(\widehat{\btheta}_c) \right\}_{\boeta_2} \widetilde{\bz}_{i2}(\mD)^\top ,
\end{align*}
are used to estimate the standard errors of $\widehat{\mu}_i(\bs)$ and $\log \{ \widehat{\sigma}_i(\bs) \}$ respectively.

Following \cite{Ruppert}, tuning parameters $\lambda_1, \lambda_2$ can be selected by minimizing the generalized cross-validation statistic
\begin{align}
\mbox{GCV}(\lambda_1, \lambda_2)&=\frac{n^{-1} \sum \limits_{k=1}^K \widetilde{\bPsi}^\top_k(\widehat{\btheta}_{Vm}) \left\{ \widetilde{\bC}^{-1}(\widehat{\btheta}_{Vm}) \right\}_{k,k} \widetilde{\bPsi}_k(\widehat{\btheta}_{Vm})}{\left(1-n^{-1}\mbox{trace}\left[ \left\{ \sum \limits_{k=1}^K \widetilde{\bPi}_k(\widehat{\btheta}_{Vm}) + \blambda_{\boeta_1} + \blambda_{\boeta_2} \right\}^{-1} \sum \limits_{k=1}^K \widetilde{\bPi}_k(\widehat{\btheta}_{Vm}) \right] \right)^2}.
\label{e:GCV}
\end{align}
Due to the heterogeneous nature of the regression coefficients $\boeta_1, \boeta_2$, the term $\blambda_{\boeta_1} + \blambda_{\boeta_2}$ induces smoothing of $\widehat{\mu}_i(\bs)$ and $\widehat{\sigma}_i(\bs)$ only within each subset $\mD_k$. As a result, these may exhibit discontinuities at the boundaries of the subsets $\mD_k$. These discontinuities may be desirable; see for example \cite{Kim-etal}. This phenomenon was also pointed out by \cite{Heaton-etal-2019} for the local approximate Gaussian process. The authors note that these discontinuities are typically small enough so as to be undetectable in visual representations, a phenomenon we have also observed. If spatial smoothness of $\widehat{\mu}_i(\bs)$ and $\widehat{\sigma}_i(\bs)$ is critical, spatial interpolation may be used in post-processing.

\section{Simulations}\label{s:simulations}

We investigate the finite sample performance of the proposed meta-estimator $(\widehat{\alpha}_m, \widehat{\phi}_m, \widehat{\bbeta}^\top_m)^\top$ derived from equation \eqref{e:meta} and $(\widehat{\alpha}_{Vm}, \widehat{\phi}_{Vm}, \widehat{\boeta}^\top_{Vm}, \widehat{\bbeta}^\top_{3Vm})^\top$ derived from equation \eqref{e:meta-VCM}. Throughout, $\mD$ consists of a square grid of evenly spaced locations. The Brown-Resnick processes $\{\mX\}_{i=1}^n$ are independently simulated using the \verb|SpatialExtremes| \citep{SpatialExtremes} R package with unit Fr\'{e}chet margins and values of $\alpha$ and $\phi$ specified below. Then $\{\mY_i \}_{i=1}^n$ are computed following the relationship in equation \eqref{e:Y-X-rel} with values of $\mu_i(\bs)$, $\sigma_i(\bs)$ and $\xi_i(\bs)$ specified below. All simulations are run on a standard Linux cluster with CCL analyses performed in parallel across $K$ CPUs with 1GB of RAM.

In the first set of simulations, we consider a $d=400$-dimensional square spatial domain $\mD=\{\bs_j \}_{j=1}^{400}$, $\bs_j \in \mathbb{R}^2$, with $n=1000$ and consider a simple model from Section \ref{ss:MSP:model}:
\begin{align*}
\mu_i(\bs; \bbeta_1)&=\bs^\top \bbeta_1,\quad 
\sigma_i(\bs; \bbeta_2)=\exp(\beta_2),\quad
\xi_i(\bs; \bbeta_3)= \beta_3,
\end{align*}
with $\btheta=(\alpha, \phi, \bbeta_1, \beta_2, \beta_3)^\top \in \mathbb{R}^6$, $\bbeta_1 = (\beta_{11}, \beta_{12})^\top$. We evaluate the performance of $\widehat{\btheta}_m$ in equation \eqref{e:meta} and its covariance with $\bJ^{-1}(\widehat{\btheta}_c)$, with $\mD$ evenly partitioned based on nearest locations into $K=16,10,8,1$ square regions of size $d_k=25,40,50,400$, $k=1, \ldots, K$, respectively (see Figure \ref{f:ex_partitions} for $K=16$). Clearly, $K=1$ corresponds to the traditional CCL analysis with no partitioning of the spatial domain. We consider three settings: in Setting I, we set threshold $u_j$ to the 80\% quantile of $\{y_i(\bs_j)\}_{i=1}^n$ and true parameter values to $\btheta_0=(0.8, 10, 0.5, 0.5, 1.5, 0.2)^\top$; in Setting II, we set $u_j$ the 90\% quantile and $\btheta_0=(1,10,0.5,0.5,1.5,0.2)^\top$; in Setting III, we set $u_j$ the 90\% quantile and $\btheta_0=(1,10,0.5,0.5,1.5,-0.2)^\top$. We report the asymptotic standard error (ASE), bias (BIAS) and 95\% confidence interval coverage (CP) averaged across 500 simulations in Table \ref{t:sim_400} for Settings I, II and III. 

The most pronounced trend in the BIAS is visible in Setting III: as $K$ decreases, the BIAS increases for all parameter estimates with the exception of $\widehat{\phi}_m$, which tends to decrease; this trend is mirrored to a lesser extent in Settings I and II. For example, in Setting I the BIAS for $\widehat{\alpha}_m$ is $-2.07$ (Monte Carlo standard error $2.2 \times 10^{-2}$) for $K=1$ compared to $0.41$ ($0.95 \times 10^{-2}$) for $K=16$; on the other hand, the BIAS for $\widehat{\phi}_m$ is $8.5 \times 10^{-2}$ (Monte Carlo standard error $1.1$) for $K=1$ compared to $9.4 \times 10^{-2}$ ($1.1$) for $K=16$. The observed trend in $\widehat{\alpha}_m, \widehat{\beta}_{11,m}, \widehat{\beta}_{2,m}, \widehat{\beta}_{3,m}$ corroborates existing literature on the increase in bias of the pairwise CL as the number of observation locations increases. The observed trend in $\widehat{\phi}_m$ is not surprising since estimation of the range is easier with more observation locations at a range of distances. Confidence interval coverage is appropriate for all settings. Across all settings, the ASE tends to increase as $K$ decreases for all parameter estimates except $\widehat{\bbeta}_1$. This reflects the fact that estimation of location parameters $\bbeta_1$ is easiest. Moreover, likelihood-based estimation with negative shape $\xi(\bs)$ is notoriously difficult and, broadly speaking, undefined for $\xi(\bs)<-0.5$ \citep{Smith-1985, Padoan-Ribatet-Sisson}. The performance of our method is therefore surprisingly good in Setting III. Mean elapsed times are reported
\begin{table}[H]
\centering
\caption{Simulation metrics for Settings I, II and III in the first set of simulations. \label{t:sim_400}}
\subfloat[Setting I: $u_j$ the 80\% quantile, $\btheta_0=(0.8, 10, 0.5, 0.5, 1.5, 0.2)^\top$.]{
\ra{0.8}
\begin{tabular}{lrrrrrrr}
Metric & $K$ & $\alpha$ & $\phi$ & $\beta_{1,1}$ & $\beta_{1,2}$ & $\beta_2$ & $\beta_3$ \\
\multirow{4}{*}{$\mbox{BIAS} \times 10^{-3}$} & $16$ & $0.41$ & $94.19$ & $-1.82$ & $-2.97$ & $4.76$ & $0.26$ \\ 
& $10$ & $-0.36$ & $110.48$ & $-1.53$ & $-3.27$ & $4.76$ & $0.09$ \\ 
& $8$ & $-0.88$ & $120.11$ & $-1.76$ & $-3.56$ & $5.65$ & $-0.34$ \\ 
& $1$ & $-2.07$ & $85.12$ & $-1.62$ & $-2.56$ & $4.07$ & $-0.86$ \\ 
\multirow{4}{*}{ASE$\times 10^{-2}$} & $16$ & $0.93$ & $107.59$ & $1.63$ & $1.63$ & $4.65$ & $2.31$ \\ 
& $10$ & $1.49$ & $111.66$ & $1.60$ & $1.61$ & $4.71$ & $2.36$ \\ 
& $8$ & $1.43$ & $110.91$ & $1.55$ & $1.57$ & $4.69$ & $2.36$ \\ 
& $1$ & $2.14$ & $113.47$ & $1.56$ & $1.56$ & $4.96$ & $2.56$ \\   
\multirow{4}{*}{CP} & $16$ & $0.95$ & $0.96$ & $0.95$ & $0.95$ & $0.94$ & $0.93$ \\ 
& $10$ & $0.94$ & $0.95$ & $0.94$ & $0.94$ & $0.95$ & $0.94$ \\ 
& $8$ & $0.93$ & $0.95$ & $0.95$ & $0.94$ & $0.95$ & $0.94$ \\ 
& $1$ & $0.93$ & $0.96$ & $0.96$ & $0.94$ & $0.95$ & $0.94$ \\ 
\end{tabular}
}\\
\subfloat[Setting II: $u_j$ the 90\% quantile, $\btheta_0=(1,10,0.5,0.5,1.5,0.2)^\top$.]{
\ra{0.8}
\begin{tabular}{lrrrrrrr}
Metric & $K$ & $\alpha$ & $\phi$ & $\beta_{1,1}$ & $\beta_{1,2}$ & $\beta_2$ & $\beta_3$\\
\multirow{4}{*}{$\mbox{BIAS} \times 10^{-3}$} & $16$ & $2.75$ & $85.88$ & $-5.41$ & $-6.72$ & $5.57$ & $1.81$ \\ 
& $10$ & $2.18$ & $103.13$ & $-5.18$ & $-7.87$ & $7.20$ & $1.03$ \\ 
& $8$ & $1.06$ & $120.41$ & $-5.83$ & $-8.58$ & $8.81$ & $0.47$ \\ 
& $1$ & $-1.81$ & $73.49$ & $-5.56$ & $-6.65$ & $8.14$ & $-1.42$ \\ 
\multirow{4}{*}{ASE$\times 10^{-2}$} & $16$ & $1.40$ & $117.93$ & $3.36$ & $3.36$ & $6.60$ & $3.06$ \\ 
& $10$ & $2.35$ & $123.88$ & $3.24$ & $3.28$ & $6.63$ & $3.11$ \\ 
& $8$ & $2.23$ & $122.57$ & $3.10$ & $3.13$ & $6.57$ & $3.11$ \\ 
& $1$ & $3.15$ & $125.56$ & $2.93$ & $2.94$ & $6.73$ & $3.30$ \\ 
\multirow{4}{*}{CP} & $16$ & $0.93$ & $0.94$ & $0.95$ & $0.95$ & $0.93$ & $0.93$ \\ 
& $10$ & $0.91$ & $0.95$ & $0.95$ & $0.94$ & $0.93$ & $0.94$ \\ 
& $8$ & $0.92$ & $0.95$ & $0.95$ & $0.94$ & $0.93$ & $0.93$ \\ 
& $1$ & $0.94$ & $0.94$ & $0.95$ & $0.95$ & $0.94$ & $0.94$ \\ 
\end{tabular}
}\\
\subfloat[Setting III: $u_j$ the 90\% quantile, $\btheta_0=(1,10,0.5,0.5,1.5,-0.2)^\top$.]{
\ra{0.8}
\begin{tabular}{lrrrrrrr}
Metric & $K$ & $\alpha$ & $\phi$ & $\beta_{1,1}$ & $\beta_{1,2}$ & $\beta_2$ & $\beta_3$\\
\multirow{4}{*}{$\mbox{BIAS} \times 10^{-3}$} & $16$ & $2.27$ & $24.25$ & $-1.36$ & $-1.62$ & $2.10$ & $0.46$ \\ 
& $10$ & $1.84$ & $3.79$ & $-1.29$ & $-1.89$ & $4.39$ & $-0.83$ \\ 
& $8$ & $0.60$ & $15.05$ & $-1.39$ & $-2.03$ & $5.09$ & $-1.15$ \\ 
& $1$ & $-4.18$ & $-13.40$ & $-2.05$ & $-2.37$ & $7.27$ & $-3.00$ \\ 
\multirow{4}{*}{$\mbox{ASE} \times 10^{-2}$} & $16$ & $1.40$ & $99.09$ & $1.20$ & $1.20$ & $4.26$ & $1.34$ \\ 
& $10$ & $2.33$ & $109.11$ & $1.19$ & $1.21$ & $4.27$ & $1.40$ \\ 
& $8$ & $2.21$ & $107.31$ & $1.16$ & $1.17$ & $4.27$ & $1.42$ \\ 
& $1$ & $3.15$ & $116.04$ & $1.17$ & $1.17$ & $4.30$ & $1.64$ \\ \ 
\multirow{4}{*}{CP} & $16$ & $0.93$ & $0.94$ & $0.93$ & $0.94$ & $0.94$ & $0.93$ \\ 
& $10$ & $0.91$ & $0.94$ & $0.93$ & $0.94$ & $0.93$ & $0.91$ \\ 
& $8$ & $0.91$ & $0.94$ & $0.94$ & $0.94$ & $0.93$ & $0.92$ \\ 
& $1$ & $0.93$ & $0.93$ & $0.93$ & $0.95$ & $0.94$ & $0.93$ \\ 
\end{tabular}
}
\end{table}
\noindent in Table \ref{t:first:time} and highlight the significant computational gain of our partitioning approach.

\begin{table}[h]
\centering
\caption{Mean elapsed time in minutes for the first set of simulations. \label{t:first:time}}
\ra{0.8}
\begin{tabular}{lcccc}
& $K=16$ & $K=10$ & $K=8$ & $K=1$ \\
Setting I & $11.2$ & $16.1$ & $22.4$ & $873.6$  \\
Setting II & $13.7$ & $18.6$ & $23.6$ & $845.7$ \\
Setting III & $18.6$ & $24.2$ & $29.7$ & $877.1$ \\
\end{tabular}
\end{table}

In the second set of simulations, we consider a larger $d=900$-dimensional square spatial domain $\mD=\{\bs_j \}_{j=1}^{900}$, $\bs_j \in \mathbb{R}^2$, with $n=1000$ and the same model as the first set of simulations. We evaluate the performance of $\widehat{\btheta}_m$ in equation \eqref{e:meta} and its covariance with $\bJ^{-1}(\widehat{\btheta}_c)$, with $\mD$ evenly partitioned based on nearest locations into $K=36$ square regions of size $d_k=25$, $k=1, \ldots, K$. We consider two settings: in Setting I, we set $u_j$ to the 90\% quantile of $\{y_i(\bs_j)\}_{i=1}^n$; in Setting II, we set $u_j$ to the 95\% quantile. In both settings, true parameter values are set to $\btheta_0=(1,10,0.5,0.5,1.5,0.2)^\top$. We report the empirical standard error (ESE), ASE, BIAS, CP and mean 95\% confidence interval length (LEN) averaged across 500 simulations for Settings I and II in Table \ref{t:sim_900}. 
\begin{table}[H]
\centering
\caption{Simulation metrics for Settings I and II in the second set of simulations.\label{t:sim_900}}
\subfloat[Setting I: $u_j$ the 90\% quantile.]{
\ra{0.8}
\begin{tabular}{rrrrrr}
Parameter & ESE$\times 10^{-1}$ & ASE$\times 10^{-1}$ & $\mbox{BIAS} \times 10^{-2}$ & CP & LEN$\times 10^{-1}$\\ 
$\alpha$ & $0.11$ & $0.11$ & $0.11$ & $0.93$ & $0.42$ \\ 
$\phi$ & $11.00$ & $11.00$ & $5.40$ & $0.94$ & $41.00$ \\ 
$\beta_{1,1}$ & $0.25$ & $0.25$ & $-0.33$ & $0.96$ & $0.98$ \\ 
$\beta_{1,2}$ & $0.24$ & $0.25$ & $-0.09$ & $0.94$ & $0.98$ \\ 
$\beta_2$ & $0.68$ & $0.63$ & $-0.17$ & $0.93$ & $2.50$ \\ 
$\beta_3$ & $0.31$ & $0.27$ & $0.25$ & $0.92$ & $1.10$ \\ 
\end{tabular}
}\\
\subfloat[Setting II: $u_j$ the 95\% quantile.]{
\ra{0.8}
\begin{tabular}{rrrrrr}
Parameter & ESE$\times 10^{-1}$ & ASE$\times 10^{-1}$ & $\mbox{BIAS} \times 10^{-2}$ & CP & LEN$\times 10^{-1}$ \\ 
$\alpha$ & $0.18$ & $0.16$ & $0.15$ & $0.92$ & $0.64$ \\ 
$\phi$ & $16.00$ & $15.00$ & $17.00$ & $0.94$ & $58.00$ \\ 
$\beta_{1,1}$ & $0.50$ & $0.50$ & $-0.32$ & $0.96$ & $2.00$ \\ 
$\beta_{1,2}$ & $0.48$ & $0.50$ & $0.29$ & $0.96$ & $2.00$ \\ 
$\beta_2$ & $1.10$ & $0.94$ & $-2.30$ & $0.92$ & $3.70$ \\ 
$\beta_3$ & $0.42$ & $0.34$ & $0.91$ & $0.91$ & $1.40$ \\
\end{tabular}
}
\end{table}

The ASE of $\widehat{\btheta}_m$ approximates the ESE, supporting the use of the asymptotic covariance formula in Theorem \ref{t:meta-asy} in finite samples. For example, the ESE and ASE for $\widehat{\beta}_{3,m}$ are $3.1\times 10^{-2}$ and $2.7 \times 10^{-2}$ (Monte Carlo standard error $5.1\times 10^{-3}$) respectively in Setting I compared to $4.2\times 10^{-2}$ and $3.4 \times 10^{-2}$ ($7.8 \times 10^{-3}$) respectively in Setting II. Additionally, the bias of $\widehat{\btheta}_m$ is negligible. We observe appropriate $95$\% confidence interval coverage. Generally, the performance of $\widehat{\btheta}_m$ and its estimated covariance are poorer in Setting II than in Setting I, with larger ESE, ASE, absolute BIAS and LEN. This is explained by the larger quantile in Setting II, which essentially reduces the amount of information used by the CCL. Mean elapsed times are $28$ and $42$ minutes for Settings I and II respectively. The mean elapsed time of $28$ minutes in Setting I is only slightly longer than the mean elapsed time of $13.7$ minutes for $K=16$ in Setting II of the first set of simulations. The latter simulation fits a comparable model to Setting I in each spatial subset since the block size $d_k$, true value $\btheta_0$ and quantile $u_j$ are the same. The slightly longer elapsed time in Setting I is due to obtaining starting values at a greater number of locations, inversion of a larger matrix $\bC(\widehat{\btheta}_c) \in \mathbb{R}^{K(2+p)\times K(2+p)}$ and the fact that the elapsed time of the total procedure depends on the slowest elapsed time of the $K$ CCL analyses. Nonetheless, the computation time for Setting I is comparable to the computation time of $30$ minutes for $K=8$ in Setting II of the first set of simulations. This comparison highlights the scalability of our partitioning approach.

In the third set of simulations, we consider a $d=400$-dimensional square spatial domain $\mD=\{\bs_j\}_{j=1}^{400}$, $\bs_j=(s_{1,j},s_{2,j}) \in \mathbb{R}^2$, with $n=2000$ and a spatially-varying coefficient model from Section \ref{ss:SVCM:model} with $z_{i1}(\bs)=z_{i2}(\bs)=1$ (i.e., $t=1$ and subscript $t$ is omitted): $\mu_i(\bs)=b_1(\bs)$, $\sigma_i(\bs)=\exp(b_2(\bs))$, and $\xi_i(\bs; \bbeta_3)= \beta_3$. We consider two settings for $b_1(\bs)$ and $b_2(\bs)$, $\bs=(s_1,s_2)$. In Setting I, $b_1(\bs)=(s_1^4 + s_2^4 + s_1 s_2)/d^2$ and $b_2(\bs)=(s^2_1+s^2_2)^{1/2}/10$. In Setting II, we let $b_1(\bs)$ and $b_2(\bs)$ be random draws from a Gaussian random field with Mat\'{e}rn covariance structure: two observations $b_j(\bs_1)$ and $b_j(\bs_2)$, $j\in \{1,2\}$, separated by a Euclidean distance of $t$ have covariance $t^5 K_5 (t)/\{2^4\Gamma(5) \}$, where $\Gamma$ is the Gamma function and $K_{5}$ is the modified Bessel function of the second kind. In both settings, we partition $\mD$ evenly based on nearest locations into $K=16$ square regions of size $d_k=25$, $k=1, \ldots, K$ (see Figure \ref{f:ex_partitions}) and approximate $b_1(\bs)$ and $b_2(\bs)$ using the same basis function expansion as follows. In each subset $k \in \{1, \ldots, K\}$, we specify knot locations $\{\boldsymbol{\kappa}_{jk}\}_{j=1}^{10}$ at 10 locations chosen by minimizing a geometric space-filling criterion \citep{Royle-Nychka}. We approximate $b_1(\bs)$ and $b_2(\bs)$ by linear combinations of Gaussian radial spline basis functions $\{C(||\bs-\boldsymbol{\kappa}_{jk}||)\}_{j=1}^{10}$, $\bs \in \mD_k$, where $C(0)=1$, $C(d)=
\exp ( -0.05d^2)$ for $d>0$. Formally, we let $\phi_{j,k}(\bs) \equiv \phi_j(\bs)$ and for $\bs=(s_1,s_2) \in \mD_k$, we define $\phi_1(\bs)=1$, $\phi_2(\bs)=s_1$, $\phi_3(\bs)=s_2$, $\phi_j(\bs)=C(||\bs-\boldsymbol{\kappa}_{j-3k}||)$, $j=4, \ldots, 13$, and $\widetilde{z}_{i1,j}(\bs)=\widetilde{z}_{i2,j}(\bs)=\phi_j(\bs)$. Finally, we approximate 
\begin{align*}
b_1(\bs)\approx \sum_{j=1}^{13} \widetilde{z}_{i1,j}(\bs) \eta_{1,j,k},\quad b_2(\bs)\approx \sum_{j=1}^{13} \widetilde{z}_{i2,j}(\bs) \eta_{2,j,k}, \quad \bs \in \mD_k,
\end{align*}
for some unknown parameters $\{\eta_{1,j,k}, \eta_{2,j,k}\}_{j=1}^{13}$. We estimate $\btheta=(\alpha, \phi, \boeta^\top, \beta_3)^\top$ with $\widehat{\btheta}_{Vm}$ in equation \eqref{e:meta-VCM} and its covariance with $\bJ^{-1}(\widehat{\btheta}_c)$ using equation \eqref{e:Vm-dist}, with $\lambda_1, \lambda_2 \in \{ 0, 0.05, 0.1\}$. The partition of $\mD$ with $K=16$ gives $\btheta \in \mathbb{R}^{419}$, where $419=2+13\times K+q_3$. We set $u_j$ to the 90\% quantile of $\{y_i(\bs_j)\}_{i=1}^n$. True parameter values of $\alpha, \phi, \beta_3$ are set to $1, 10, 0.2$ respectively. Define the absolute error deviation, its average and its maximum as
\begin{equation*}
\begin{split}
\mbox{AED}_j(\bs) = \left| \widehat{b}_j(\bs) - b_j(\bs) \right| / \left[ \max\left\{ b_j(\mD) \right\} - \min\left\{ b_j(\mD) \right\} \right],~~~~~\\
\mbox{aAED}_j=\sum_{\bs \in \mD} \mbox{AED}_j(\bs)/d, \quad \mbox{mAED}_j=\max_{\bs \in \mD} \mbox{AED}_j(\bs), \quad j=1,2,
\end{split}
\end{equation*}
respectively. We report the BIAS, ASE and CP for estimates of $\alpha,\phi,\bbeta_3$ and the $\mbox{aAED}_j$, $\mbox{mAED}_j$ and $\mbox{CP}_j$ of $b_j(\bs)$, $j=1,2$, with optimal $\lambda_1, \lambda_2$ selected using the generalized cross-validation statistic in equation \eqref{e:GCV}, averaged across 100 simulations for Settings I and II in Table \ref{t:sim_VCM}. Selected values of $\lambda_1,\lambda_2$ across the 100 simulations are reported in the online supplementary materials. 
\begin{table}[H]
\centering
\caption{Varying coefficient model simulation results for $n=2000$, $d=400$, $K=16$ square regions of size $d_k=25$, $k=1, \ldots, K$. Simulation metrics for $b_1(\bs)$ and $b_2(\bs)$ are averaged over observation locations. \label{t:sim_VCM}}
\subfloat[Setting I: $b_1(\bs)=(s_1^4 + s_2^4 + s_1 s_2)/d^2$, $b_2(\bs)=(s^2_1+s^2_2)^{1/2}/10$. ]{
\ra{0.8}
\begin{tabular}{rrrrrr}
Parameter & $\mbox{BIAS} \times 10^{-2}$ & ASE$\times 10^{-1}$ & aAED & mAED& CP \\ 
$\alpha$ & $-0.40$ & $0.08$ & -- & -- & $0.93$ \\ 
$\phi$ & $2.40$ & $7.10$ & -- & -- & $0.96$ \\ 
$b_1(\bs)$ & -- & -- & $2.40$ & $8.80$ & $0.94$  \\
$b_2(\bs)$ & -- & -- & $4.00$ & $9.60$ & $0.92$ \\
$\beta_3$ & $-0.86$ & $0.18$ & -- & -- & $0.86$ \\ 
\end{tabular}
}\\
\subfloat[Setting II: $b_1(\bs), b_2(\bs)$ drawn from Gaussian random fields.]{
\ra{0.8}
\begin{tabular}{rrrrrr}
Parameter & $\mbox{BIAS} \times 10^{-2}$ & ASE$\times 10^{-1}$ & aAED & mAED& CP \\ 
$\alpha$ & $-0.75$ & $0.08$ & -- & -- & $0.81$ \\ 
$\phi$ & $2.30$ & $7.10$ & -- & -- & $0.96$ \\ 
$b_1(\bs)$ & -- & -- & $0.36$ & $4.30$ & $0.90$ \\
$b_2(\bs)$ & -- & -- & $0.11$ & $0.38$ & $0.91$ \\
$\beta_3$ & $-0.53$ & $0.18$ & -- & -- & $0.91$ \\ 
\end{tabular}
}
\end{table}

\begin{figure*}[h]
\caption{Varying coefficient model simulation result for first simulated dataset in Setting II with $b_1(\bs), b_2(\bs)$ drawn from Gaussian random fields with Mat\'{e}rn covariance structure.\label{f:VCM-figure}}
\includegraphics[width=\textwidth]{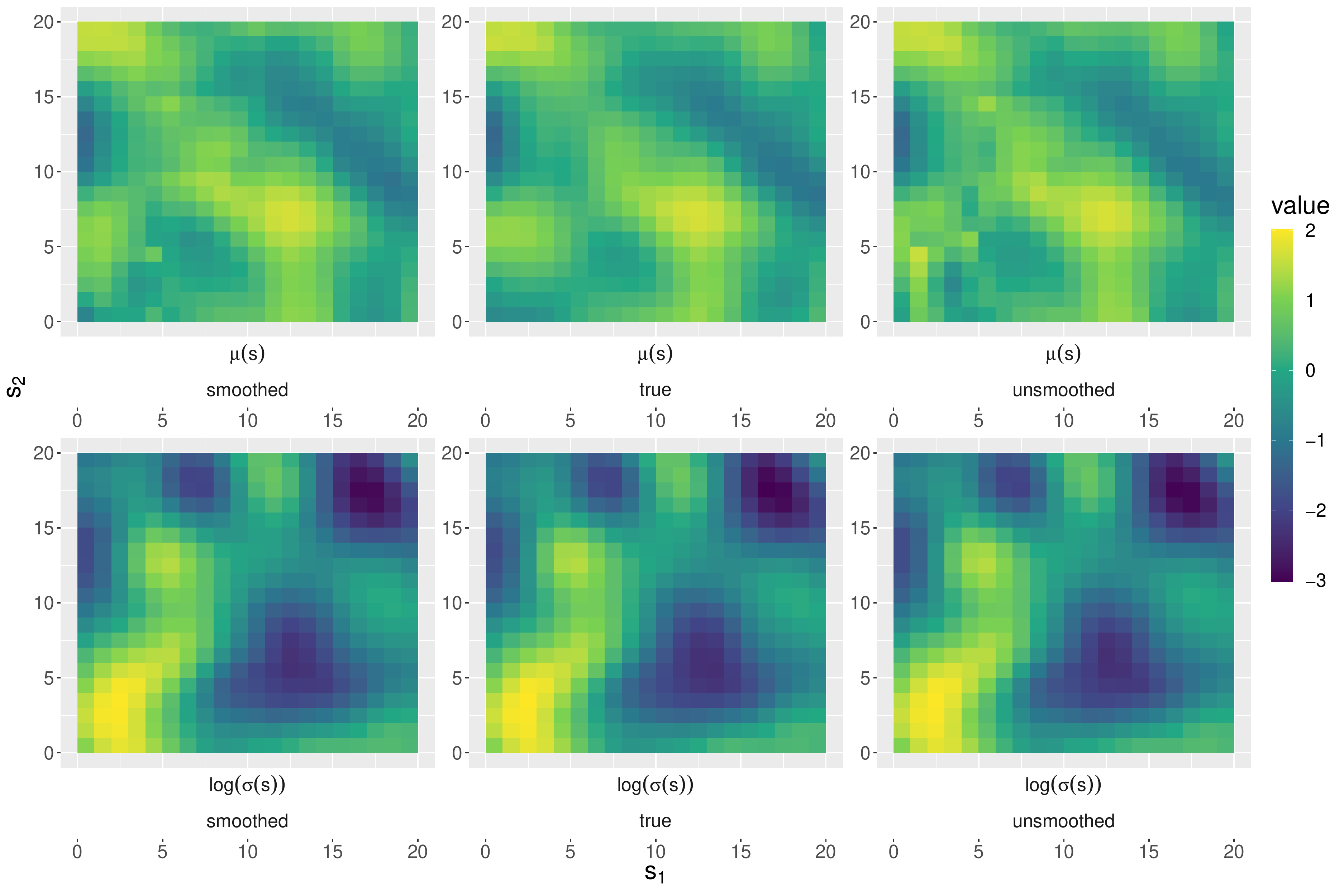}
\end{figure*}

In Setting I, BIAS, aAED and mAED are appropriately small to suggest good point estimation, and CP of estimates of $\alpha, \phi, b_1(\bs)$ and $b_2(\bs)$ show appropriate coverage of the 95\% confidence intervals. In Setting I, the $\beta_3$ parameter for the shape is disappointingly undercovered with CP of $86\%$. In Setting II, BIAS, aAED and mAED again suggest good point estimation, and we observe a slight undercoverage of parameters. The undercoverage observed in Settings I and II is potentially due to several factors. The performance of the GMM is known to deteriorate as the dimension of the estimating function $\widetilde{\bPsi}_{all}(\btheta)$ increases relative to $n$. In both settings, $\widetilde{\bPsi}_{all}(\btheta) \in \mathbb{R}^{464}$, where $464=\sum_{k=1}^K (2+p_k+q_3) \approx n/4$. Thus, $\widetilde{\bC}(\widehat{\btheta}_c)$ may yield a poor estimate of the covariance of $\widetilde{\bPsi}_{all}(\btheta)$, affecting the estimation of the covariance of $\widehat{\btheta}_{Vm}$ with $\bJ^{-1}(\widehat{\btheta}_c)$. As discussed in the first set of simulations, estimation of the shape parameter can be difficult because the bounds of the parameter space depend on observed values of $\mathcal{Y}$. This may explain the undercoverage of $\beta_3$ in Settings I and II. Finally, undercoverage of the estimate of $\alpha$ in Setting II may be due to the roughness of the location and scale parameters when simulated from the Gaussian process, which may confound the smoothness of the spatial dependence. Given the difficulty of estimating $b_1(\bs)$ and $b_2(\bs)$ in Setting II, the CP for these two functional parameters is surprisingly good. Mean elapsed time, including cross-validation over the grid of $(\lambda_1, \lambda_2)$ values, is 7.1 and 9 hours in Settings I and II respectively. Plots of the true and estimated $\mu(\bs), \sigma(\bs)$ are shown for the first simulated dataset of Setting II in Figure \ref{f:VCM-figure}: we observe slight discontinuity between spatial blocks in the estimated location $\widehat{\mu}(\bs)$, as discussed in Section \ref{ss:SVCM:integration}.

\section{Analysis of Extreme Streamflow Across the US}\label{s:data}

\begin{figure}[H]
\centering
\includegraphics[width=0.8\textwidth]{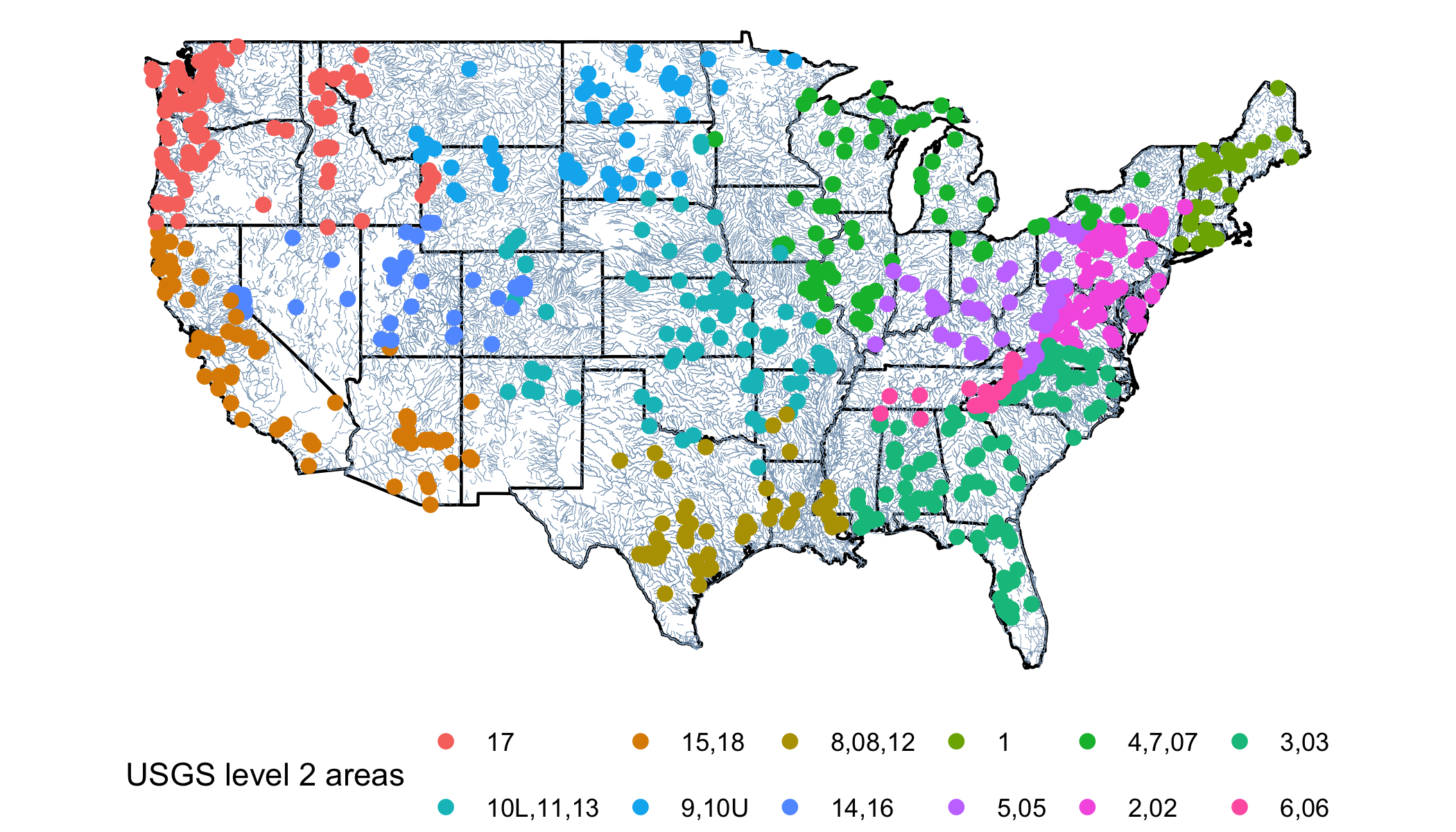}
\caption{Partitioning of 702 spatial locations into $K=12$ blocks based on USGS watershed boundary regions level 2 hydrologic unit codes (HUC02).}
\label{f:blocks}
\end{figure}

\begin{figure}[H]
\centering
\includegraphics[width=\textwidth]{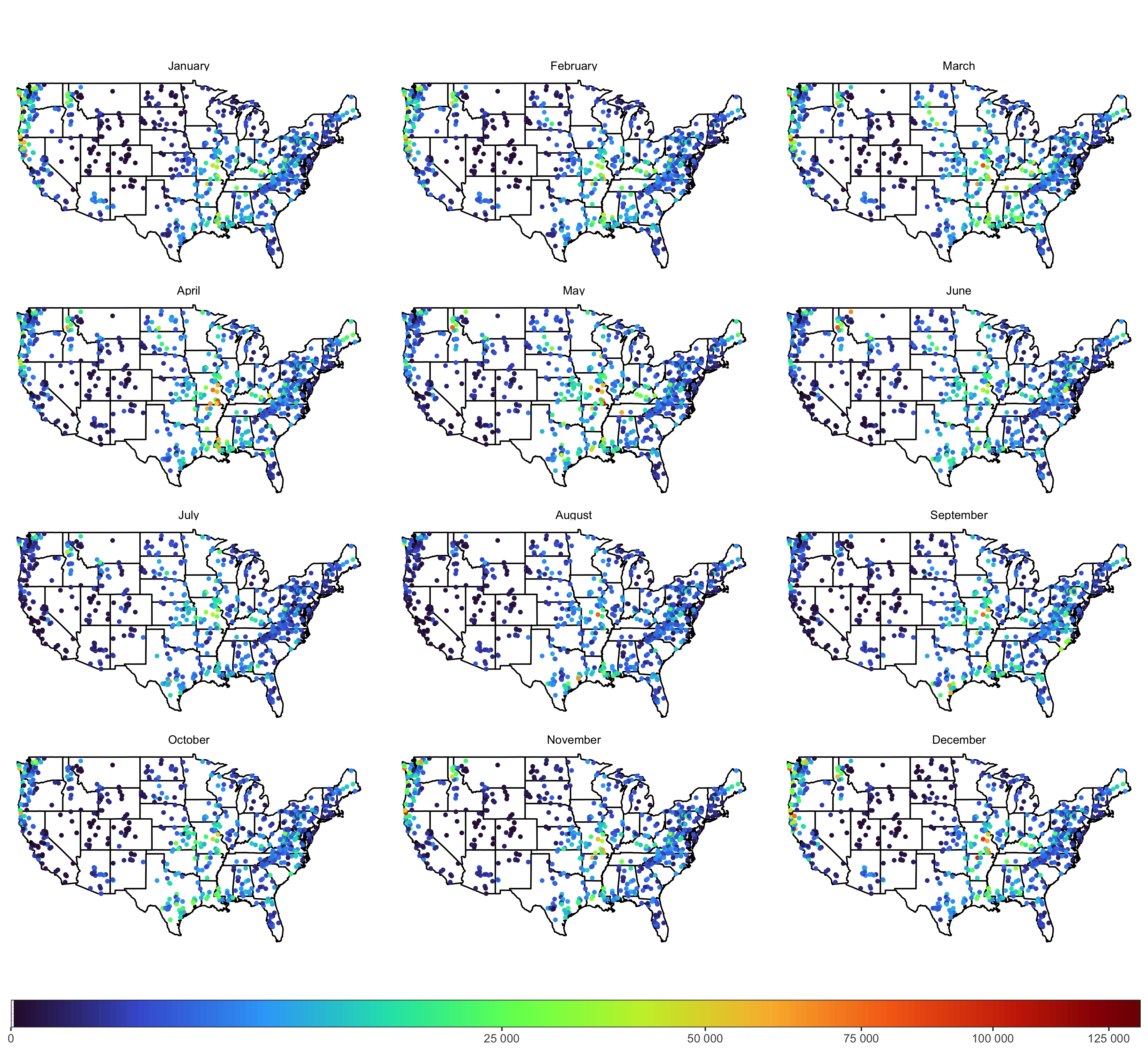}
\caption{Streamflow (cfs) 99\% sample quantiles for each month and site pooling across years.}
\label{f:bymonth-extremes}
\end{figure}

To illustrate the proposed method, we analyze monthly measurements of streamflow from 1950-2020 at 702 locations across the US as shown in Figure \ref{f:blocks}.  These locations are part of the USGS Hydro-Climatic Data Network 2009 \citep{lins2012usgs} and are selected because of their long record and because they are relatively unaffected by human activities.  The locations are partitioned into $K=12$ blocks based on the USGS watershed boundary regions (Figure \ref{f:blocks}). The response for month $i$ at location $\bs$, $y_i(\bs)$, is the monthly maximum of the daily streamflow measurements. Streamflow has strong seasonality (Figure \ref{f:bymonth-extremes}) and so we take covariates for the GEV location ($z_{i1}(\bs)$) and log scale ($z_{i2}(\bs)$) to include an intercept and four Fourier basis functions (two sine and two cosine) of the observation month to capture seasonality.  The effects of these covariates are allowed to vary spatially following Section \ref{s:SVCM}. The GEV shape parameter $\xi$ is assumed to be constant across space and time. 

While the data are block maxima, the block size of a month may be insufficient to assume the data follow a max-stable process.  Therefore, we analyze threshold exceedances using the censored likelihood in equation (\ref{e:CCL}).  To account for local heterogeneity we standardize the data at each site by subtracting the site's sample median and dividing by the difference of the site's 95\% and 5\% quantiles; all plots are made on the original data scale. We use
\noindent roughly one spatial basis function per twenty locations in each block, giving between 2 and 5 basis functions per block. The basis functions are the same Gaussian kernel functions as in Section \ref{s:simulations}. We take the threshold $u(\bs)$ at a location $\bs$ to be the level $q$ sample quantile of the observations at the location. We fit the spatial model for several $q$ and compare the results. Because many sites have a large number of zeros, we consider only $q\in\{85\%, 90\%, 95\%\}$. As shown in the online supplementary materials, the fitted values and goodness-of-fit diagnostics are similar for all three thresholds so we present results for $q=85\%$. For $q=85\%$, tuning parameters $\lambda_1=0.001, \lambda_2= 0.080$ are selected from the ranges $\lambda_1\in[0,0.003], \lambda_2\in[0,0.15]$ via the generalized cross-validation statistic in \eqref{e:GCV}. 

\begin{figure}[H]
\centering
\includegraphics[width=0.5\textwidth]{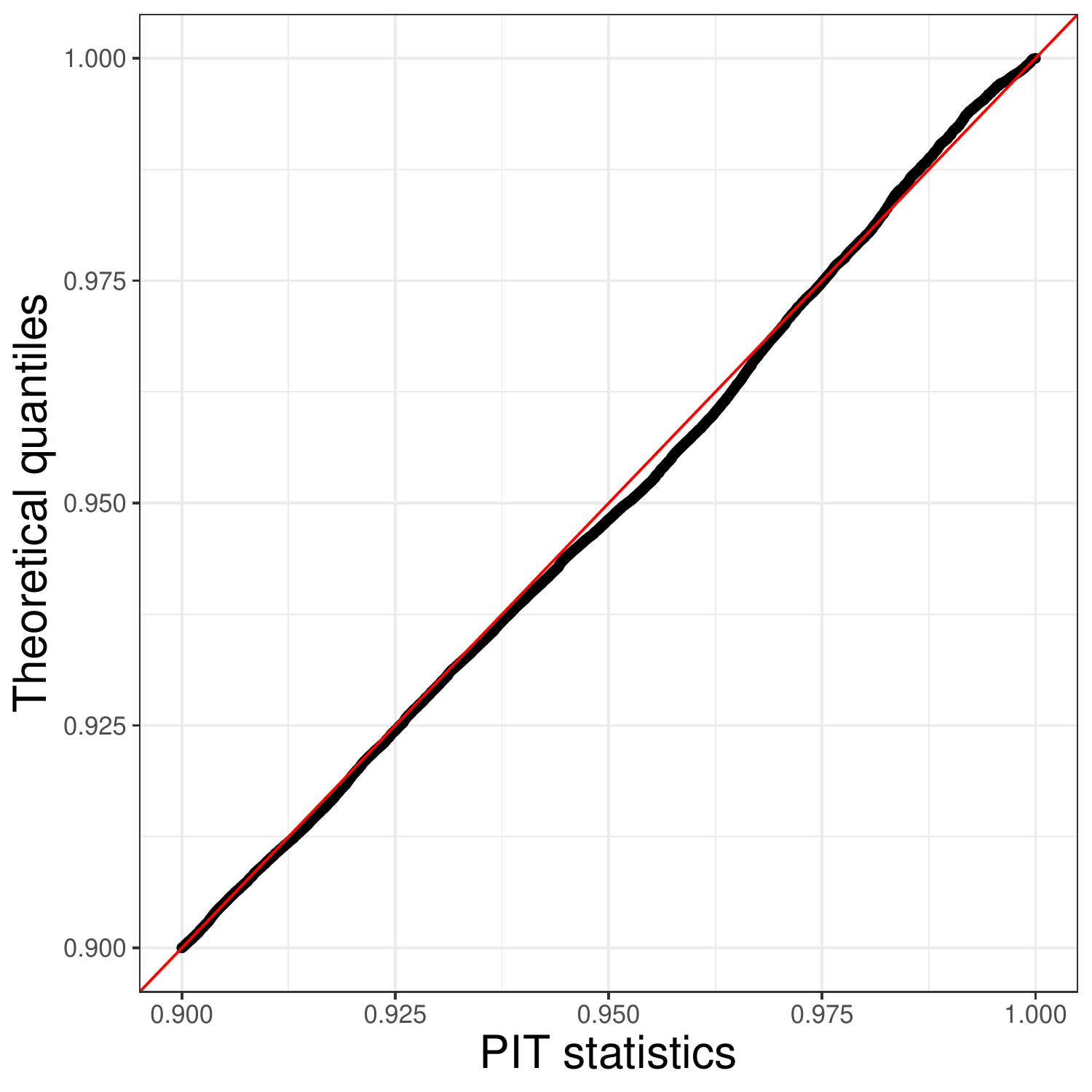}
\caption{Probability integral transform plot for model with threshold set to the $q=85\%$ quantile.}
\label{f:PIT}
\end{figure}

We use a probability integral transform plot to evaluate the fit of the model with $q=85\%$.  For each observation we compute $U_i(\bs) = \hat{F}_i\{y_i(\bs);\bs\}$, where ${\hat F}_i(y;\bs)$ is the fitted marginal GEV distribution function at site $\bs$ and time $i$.  Assuming the model fits well, the distribution of the $U_i(\bs)$ should be approximately Uniform(0,1). Figure \ref{f:PIT} shows that this is case for the fitted model.  

\begin{figure}[H]
\centering
\includegraphics[width=\textwidth]{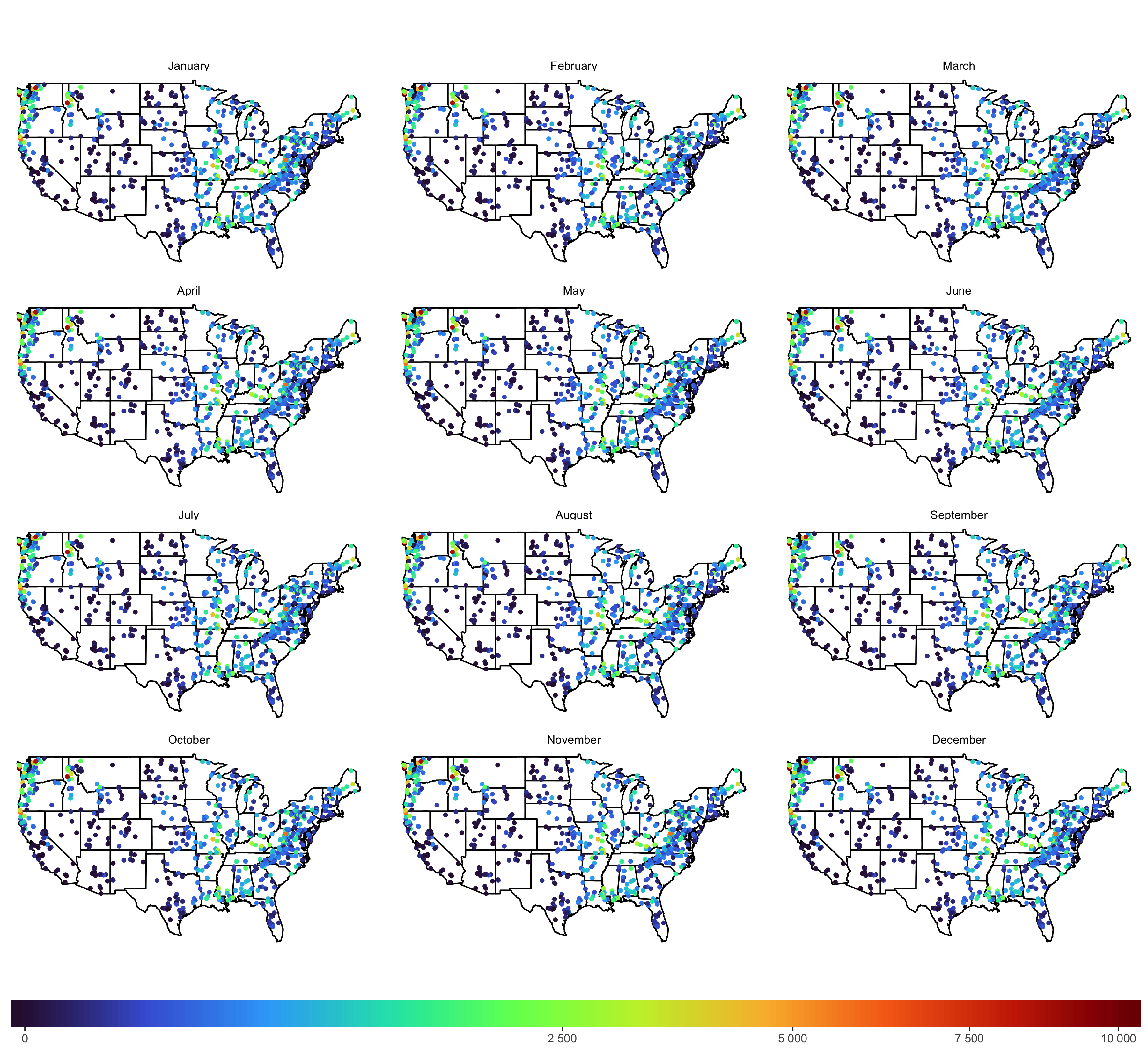}
\caption{Estimated location parameter $\widehat{\mu}_i(\bs)$ by month and site with optimal smoothing parameters and threshold set to the 85\% quantile.}
\label{f:VCM-mu}
\end{figure}

\begin{figure}[H]
\centering
\includegraphics[width=\textwidth]{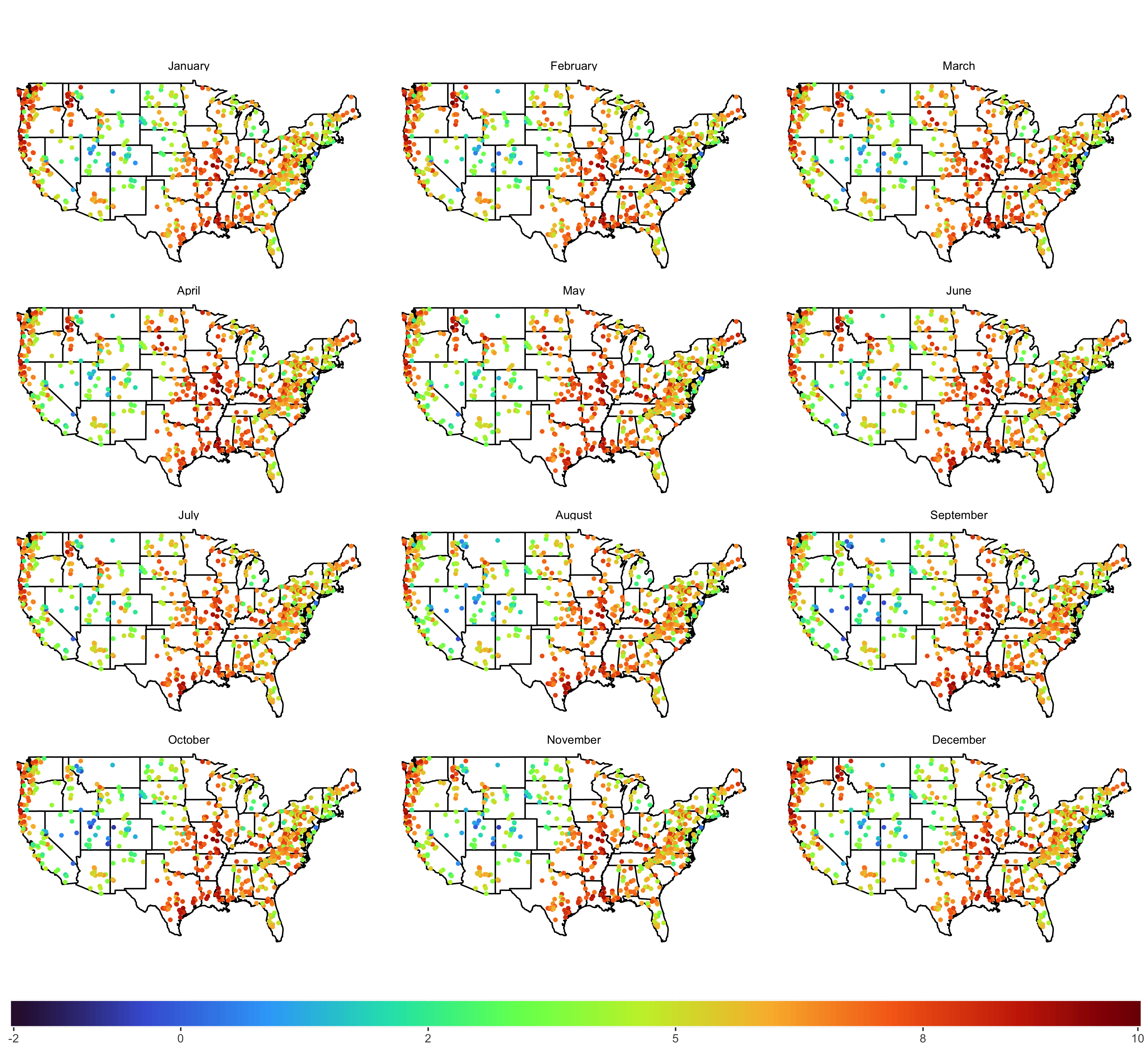}
\caption{Estimated log scale parameter $\log \{ \widehat{\sigma}_i(\bs) \}$ by month and site with optimal smoothing parameters and threshold set to the 85\% quantile.}
\label{f:VCM-logsigma}
\end{figure}

\begin{figure}[H]
\centering
\includegraphics[height=0.95\textheight]{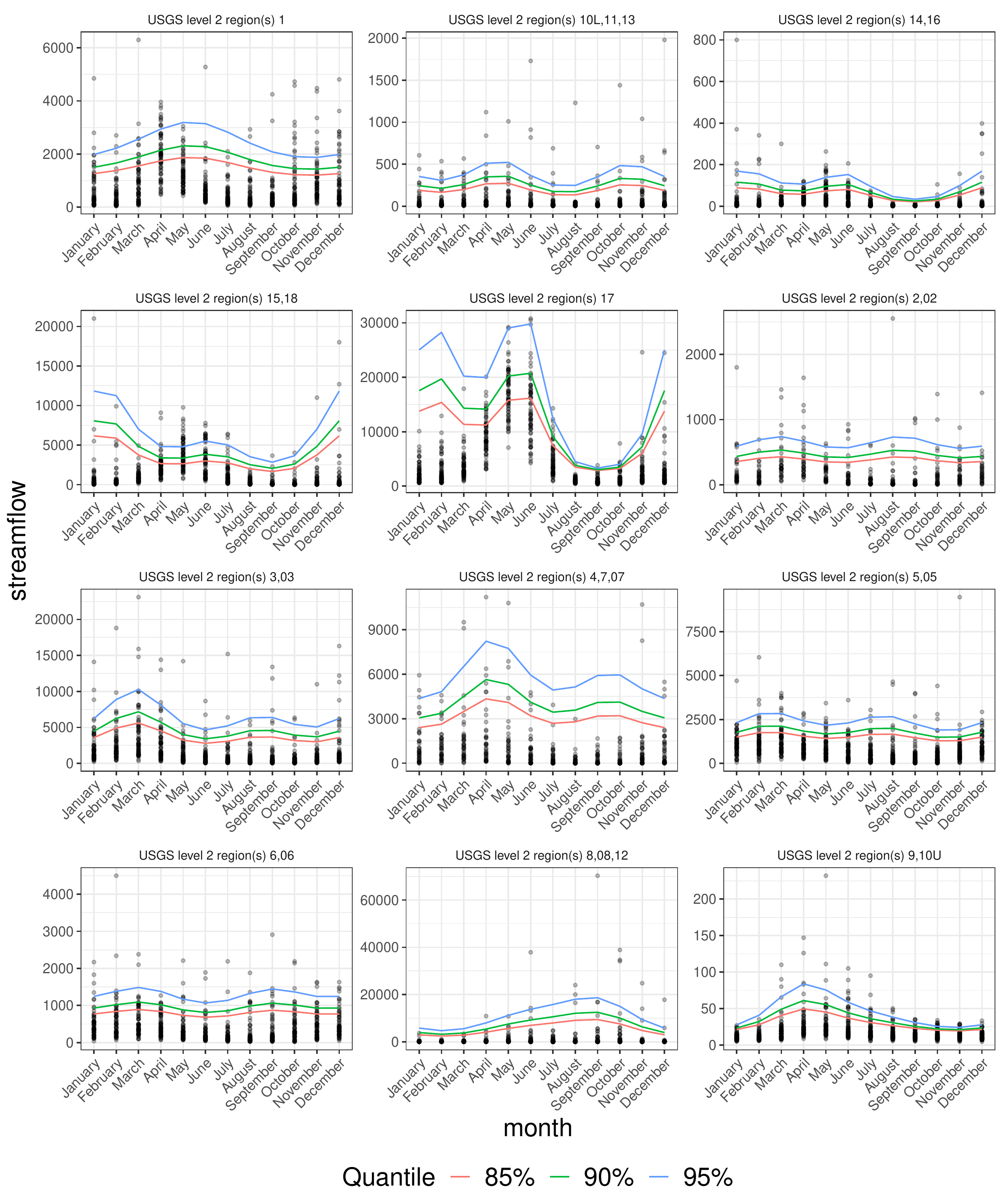}
\caption{Streamflow (cfs) observations (dots) versus fitted quantiles (lines) for one randomly-selected station in each of the $K=12$ blocks with threshold set to the $q=85\%$ quantile.}
\label{f:qvo85}
\end{figure}

Figures \ref{f:VCM-mu} and \ref{f:VCM-logsigma} map the estimated values of $\mu_i(\bs)$ and $\sigma_i(\bs)$. The location parameter varies considerably over space, with highest values in the Pacific Northwest and Missouri. The scale varies more by season, most notably in the Northern Plains.  The GEV shape parameter (standard error) is estimated to be $\widehat{\xi}_{Vm}=0.31$ ($0.0049$) giving a right-skewed distribution.  The estimated spatial dependence parameters are $\widehat{\alpha}_{Vm}=0.71$ ($0.0051$) and $\widehat{\phi}_{Vm}=0.56$ ($0.02$). Seasonal variation in Figures \ref{f:VCM-mu} and \ref{f:VCM-logsigma} is largely obscured by spatial variation, so Figure \ref{f:qvo85} plots the data (pooled across years) versus fitted GEV quantiles for one randomly-selected station in each of the $K$ blocks. The sites have prominent and varied seasonal patterns, illustrating the difficulty in modeling extremes over a large and heterogeneous region.  For example, streamflow peaks in the spring for USGS level 2 regions 10L, 11 and 13, fall for USGS level 2 regions 14 and 16 and winter for USGS level 2 regions 2 and 02; the fitted model generally captures these disparate trends. 

\begin{figure}[H]
\centering
\includegraphics[width=0.75\textwidth]{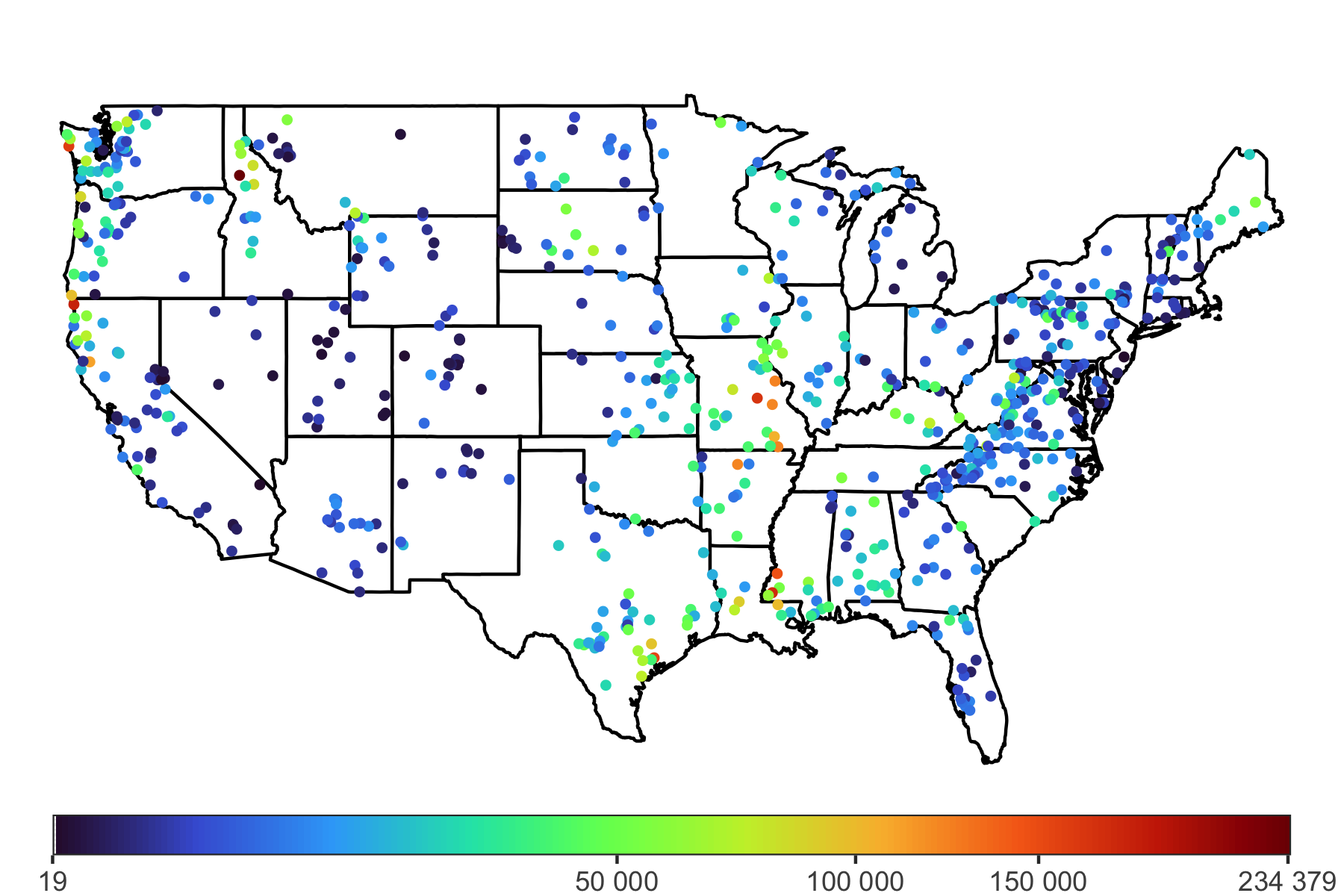}
\includegraphics[width=0.75\textwidth]{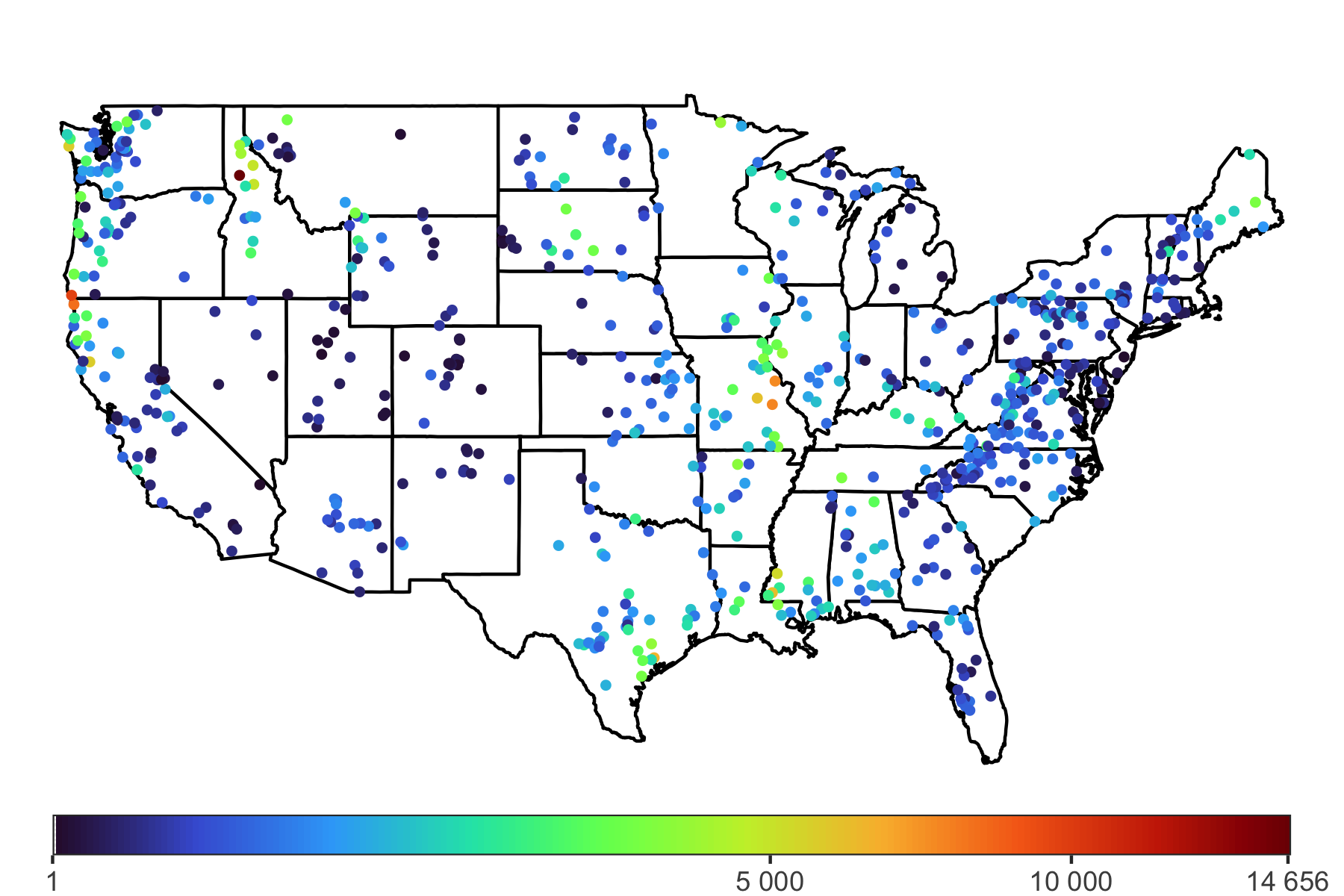}
\caption{Estimated $50$-year return levels (top) and standard errors (bottom) on the original data scale (cfs) with optimal smoothing parameters and threshold set to the $q=85\%$ quantile.}
\label{f:50y-return}
\end{figure}

While the fitted model includes seasonality, the results can also be used to estimate the distribution and return level of the annual maximum.  Let ${\hat F}_i(y; \bs)$ be the fitted GEV distribution function for month $i$ at location $\bs$, then the distribution function of the annual maximum is estimated as ${\hat F}(y;\bs) = \prod_{i=1}^{12}{\hat F}_i(y;\bs)$ (noting that the fitted GEV distribution does not change by year).  Inverting ${\hat F}(y;\bs)$ gives the estimated quantile function and thus the $r$-year return level, i.e., the $1-1/r$ quantile of the annual maximum. Figure \ref{f:50y-return} plots the estimated 50-year return level and its standard error.  The return level is maximized
\noindent for the stations at the mouth of the Mississippi River, in Southern Missouri and the Pacific Northwest.  These stations also have high sample quantiles (Figure \ref{f:bymonth-extremes}) but the fitted return levels are more stable and smooth across space. 

\section{Discussion}\label{s:discussion}

The proposed approach delivers three pivotal innovations to facilitate the efficient analysis of spatial extremes. The censored pairwise likelihood uses information from all observations, including those not theoretically extreme enough to be appropriately modelled by the MSP, for both flexible and efficient estimation. Ensuing computational difficulties are efficiently handled through a spatial partitioning approach and accompanying meta-estimator that strike a balance between optimal asymptotic efficiency and reduced finite-sample bias. Finally, the varying coefficient model formulation in the MapReduce paradigm achieves not only computationally feasible, but computationally efficient complex modeling of spatial variation in marginal parameters. We demonstrated the practicability and scalability of the proposed modeling framework through extensive simulations and the analysis of 71 years of streamflow data in the contiguous United States. The R package provided in the online supplementary materials facilitates implementation by practitioners.

The GMM suffers from well-known variance under-estimation when the sample size $n$ is small relative to the number of estimating equations $Kp$; see \cite{Hansen-Heaton-Yaron} and others in the same issue. The difficulty primarily stems from evaluating $\bC(\btheta)$ at a consistent estimator whose variability is not accounted for in $\bJ(\btheta)$ and may be high when $n$ is small. This issue is mitigated by the use of censored composite likelihood, which uses more observations than pure composite likelihood approaches. Nonetheless, the rarity of extreme events and corresponding small sample sizes may prohibit the use of the asymptotic covariance formula in Theorems \ref{t:meta-asy} and \ref{t:VCM-asy}. Two dominant strategies are available in these situations. The sample covariance $\bC(\btheta)$ may be evaluated using repeated sub-sampling schemes such as in \cite{Bai-Song-Raghu} that artificially increase the sample size. This approach, however, can be extremely computationally burdensome. Alternatively, \cite{Windmeijer} propose a finite-sample correction to $\bC(\btheta)$. Additional approaches are discussed in Chapter 8 of \cite{Hall}.

As mentioned throughout this article, the choice $K$ should be large enough that block MCCLEs are computationally fast to obtain and finite-sample bias is minimal, but small enough that a range of distances are available in each block for estimation of the dependence parameters, and in particular the range parameter $\phi$. In practice, we have found that using $d_k=25$ spatial locations per block performs well. We refer the reader to Section \ref{ss:GMM:partition} for a discussion on the choice of $K$.

Future work includes extending the methods beyond MSPs. For example, \cite{Huser-Wadsworth} propose a scale-mixture process that allows for a richer class of dependence structures than a MSP. The proposed distributed inference approach can be directly applied to any process that permits a bivariate density function and is thus broadly applicable. However, future work is needed to verify the theoretical and statistical performance of the approach for processes other than the MSP.  Future work would also extend to a spatiotemporal analysis to investigate climate change effects on the distribution of extreme streamflow. 

\appendix

\section{Appendices}

\subsection{Conditions for asymptotic results}\label{a:c}

The conditions for consistency and asymptotic normality of $\widehat{\btheta}$ can be divided into conditions for consistency and asymptotic normality of the MLCEs $\widehat{\btheta}_k$, and conditions to ensure $\widehat{\btheta}$ inherits these properties. Denote by $\Theta$ the parameter space of $\btheta$.
\begin{enumerate}[label=(C\arabic*)]
\item \label{c:1} Conditions for consistency and asymptotic normality of the MCCLEs $\{\widehat{\btheta}_k\}_{k=1}^K$, following \cite{Padoan-Ribatet-Sisson}:
\begin{enumerate}[label=(\roman*)]
\item The support of $\mathcal{Y}$ does not depend on $\btheta$.
\item The estimating function $\bPsi_{all}(\by; \btheta)$ is twice continuously differentiable with respect to $\btheta$ for all $\by \in \mY$. 
\item The expectation $E_{\btheta_0} \{ \bPsi_{all}(\btheta)\}$ has a unique zero at an merior point $\btheta_0$ of $\Theta$. 
\item The covariance matrix $\bc(\btheta) $ is finite and positive definite for all $\btheta \in \Theta$.
\item The sensitivity matrix has first derivative $ \nabla_{\btheta} \bi_k(\btheta)$ uniformly bounded for all $\btheta$ in a neighbourhood of $\btheta_0$.
\end{enumerate}
\item \label{c:2} Conditions for consistency and asymptotic normality of $\widehat{\btheta}_m$, following \cite{Hector-Song-JMLR}: for any $\delta_n \rightarrow \infty$,
\begin{align*}
\sup \limits_{\left\| \btheta - \btheta_0 \right\| \leq \delta_n} \frac{n^{1/2}}{1+n^{1/2} \left\| \btheta - \btheta_0 \right\|} \left\| \bPsi_k(\btheta) - \bPsi_k(\btheta_0) - E_{\btheta_0} \bPsi_k(\btheta) \right\| = O_p\left( n^{-1/2} \right),
\end{align*}
for all $k \in \{1, \ldots, K\}$.
\end{enumerate}
As in \cite{Padoan-Ribatet-Sisson}, asymptotic results are derived for $n\rightarrow \infty$ with fixed dimension $d$, and therefore fixed $d_k$ and $K$. Equivalent results can be derived for the asymptotic setting $d \rightarrow \infty$ due to $d_k\rightarrow \infty$ and/or $K \rightarrow \infty$. See \cite{Hector-Song-JASA, Cox-Reid, Hector-Song-JMLR} for necessary conditions in this setting.

\bibliographystyle{apalike}
\bibliography{bibliography-20201023}

\end{document}